\documentclass[a4paper,fleqn,usenatbib]{mnras}

\usepackage{newtxtext,newtxmath}

\usepackage[T1]{fontenc}
\usepackage{ae,aecompl}

\usepackage{graphicx}	
\usepackage{amsmath}	
\usepackage{amssymb}	
\usepackage{hyperref}

\def\apj{ApJ}
\def\apjs{ApJ}
\def\mnras{MNRAS}
\def\aap{A\&A}
\def\apjl{ApJ}

\def\nat{Nature}

\def\apss{Ap\&SS}

\def\aapr{Astronomy and Astrophysics Reviews}

\def\ixpe{\textit{IXPE}}
\def\xipe{\textit{XIPE}}
\def\extp{\textit{eXTP}}
\def\astrosat{\textit{AstroSat}}

\title[X-ray polarimetry timing]{An observational method for fast
  stochastic X-ray polarimetry-timing}

\author[A. Ingram \& T. Maccarone]{
Adam R. Ingram,$^{1}$\thanks{E-mail: a.r.ingram@uva.nl}
\& Thomas J. Maccarone$^{2}$
\\
$^{1}$Anton Pannekoek Institute, University of Amsterdam, Science Park 904, 1098 XH Amsterdam, The Netherlands\\
$^{2}$Department of Physics, Texas Tech University, Box 41051, Lubbock, TX 79409-1051, USA
}

\date{Accepted 2017 July 19. Received 2017 July 19; in original form 2017 June 16}

\pubyear{2017}

\begin{document}
\label{firstpage}
\pagerange{\pageref{firstpage}--\pageref{lastpage}}
\maketitle

\begin{abstract}
The upcoming launch of the first space based X-ray polarimeter in
$\sim 40$ years will provide powerful new diagnostic information to
study accreting compact objects. In particular, analysis of rapid
variability of the polarisation degree and angle will provide the
opportunity to probe the relativistic motions of material in the
strong gravitational fields close to the compact objects, and enable
new methods to measure black hole and neutron star
parameters. However, polarisation properties are measured in a
statistical sense, and a statistically significant polarisation
detection requires a fairly long exposure, even for the brightest
objects. Therefore, the sub-minute timescales of interest are not
accessible using a direct time-resolved analysis of polarisation
degree and angle. Phase-folding can be used for coherent
pulsations, but not for stochastic variability such as quasi-periodic
oscillations. Here, we introduce a Fourier method that enables
statistically robust detection of stochastic polarisation variability for
arbitrarily short variability timescales. Our method is analogous to
commonly used spectral-timing techniques. We find that it should be
possible in the near future to detect the quasi-periodic swings in
polarisation angle predicted by Lense-Thirring precession of the inner
accretion flow. This is contingent on the mean polarisation degree of
the source being greater than $\sim 4-5\%$, which is consistent with
the best current constraints on Cygnus X-1 from the late 1970s.
\end{abstract}

\begin{keywords}
methods: data analysis -- X-rays: general -- polarization -- black
hole physics
\end{keywords}



\section{Introduction}
\label{sec:intro}

Accreting compact objects radiate brightly in X-rays, enabling a view
of the region close to the horizon in the case of black holes (BHs),
or the surface in the case of neutron stars (NSs). Accretion occurs
through a geometrically thin disk, which emits a thermalised spectrum
(\citealt{Novikov1973,Shakura1973}), and a hot cloud of electrons
located close to the compact object, in which photons are Compton
up-scattered into a cut-off power-law spectrum
(\citealt{Thorne1975,Eardley1975}). The exact geometry of this
electron cloud is still debated, with candidate models including
evaporation of the inner disk into a large scale-height accretion flow
(\citealt{Eardley1975}), the base of a jet (\citealt{Markoff2005}),
and a coronal layer  held above the disk by magnetic reconnection
(\citealt{Galeev1979,Haardt1991}). In the case of  NSs, additional
radiation is associated with the surface of the compact star.

Since the region of interest closest to the compact object
cannot be directly imaged, the accretion geometry there can only be
inferred by analysing the properties of the X-ray signal. For the past
$\sim 40$ years this has been limited to analysis of the spectral and
variability properties, with particular success resulting from
combining the two disciplines
(e.g. \citealt{Miyamoto1988,Maccarone2002,Skipper2013}). Such
`spectral-timing' techniques allow, for instance, analysis of
propagating accretion rate fluctuations
(e.g. \citealt{Ingram2013,Rapisarda2017}) and reverberation mapping
(e.g. \citealt{Uttley2014}). Soon with the (proposed late 2020) launch
of the \textit{Imaging X-ray Polarimetry Explorer} (\ixpe:
\citealt{Weisskopf2016}), it will also be possible, for the first time
since \textit{OSO 8} was switched off in 1978, to observe the X-ray
polarisation of these sources. Polarimetry provides two extra
properties: the polarisation degree and angle. Analysis of these
properties as a function of energy (spectral-polarimetry) will provide
a powerful new lever arm to determine the geometry of the system and
measure parameters of the compact object
(e.g. \citealt{Stark1977,Schnittman2010,Dovciak2011}). Analysis of the
rapid  variability of the polarisation degree and angle will provide
similarly powerful diagnostics. For instance, this will allow us to
track propagation of accretion rate fluctuations from strongly to
weakly polarised regions of the accretion flow and vice versa, and
will provide a new way to disentangle scattered from directly observed
photons for the purposes of reverberation mapping.

Accreting stellar-mass BHs and NSs display a rich phenomenology of
X-ray variability properties on timescales ranging from milliseconds
to hundreds of seconds (e.g. \citealt{VDK2006}). In particular,
quasi-periodic oscillations (QPOs) are often observed. These signals
can be classified depending on the fundamental frequency of the
oscillation. Low frequency (LF) QPOs are routinely observed from both
BHs and NSs, often with a large amplitude. The observed frequency
range for BHs is $\sim 0.1-30$ Hz, with the higher frequencies
observed from NSs consistent with simple mass scaling
(\citealt{Wijnands1999,vanderKlis2005,Belloni2010}). BHs occasionally
display high frequency (HF) QPOs, with frequencies $\gtrsim 100$ Hz 
(e.g. \citealt{Morgan1997,Remillard1999,Homan2001}).
Even though these features are extremely rare and
weak, they command significant theoretical interest because their
frequencies are commensurate with the orbital frequency at the
innermost stable circular orbit (\citealt{Stella1999,Motta2014}). NSs on the
other hand display kHz QPOs which are common and often strong features
(\citealt{Strohmayer1996,vanderKlis1996}). Although it is tempting to
interpret HF QPOs as the BH equivalent of kHz QPOs, this comparison
is challenging on closer inspection (\citealt{Motta2017}).

All of these
classes of QPO are often interpreted as a geometrical effect, giving
rise to the possibility of detecting a QPO in the polarisation degree
and/or angle with a sufficiently sensitive X-ray polarimeter. In
particular, there is now mounting evidence that LF QPOs in BHs (or at
least the `Type C' subclass of LF QPOs; see
e.g. \citealt{Casella2005}) result from Lense-Thirring precession of
the inner accretion flow (\citealt{Stella1998,Ingram2009}). This is a
relativistic effect in which a spinning compact object twists up the
surrounding spacetime, inducing nodal precession in nearby orbits
inclined to the BH equatorial plane
(\citealt{Lense1918}). \cite{Ingram2016} recently discovered that the
iron K$\alpha$ fluorescence line in the spectrum of the accreting BH H
1743$-$322 rocks from red to blue shifted over the course of a QPO
cycle, confirming a distinctive prediction of the precession model
(\citealt{Schnittman2006,Ingram2012}). This model also predicts a QPO
in both the polarisation angle, resulting from the changing projected
orientation of the accretion flow, and in polarisation degree,
resulting from the expected angular dependence of Compton scattering
(\citealt{Ingram2015a}). Confirmation of these predictions would not
only provide smoking gun evidence for the precession hypothesis, but
would also provide tight geometrical constraints, particularly in
combination with QPO phase-resolved iron line modeling
(i.e. tomography: \citealt{Ingram2017}).

Detection of rapid variability of the polarisation degree and/or angle
would therefore provide a valuable probe of these systems. However,
the polarisation properties cannot be directly measured on
arbitrarily short timescales due to Poisson counting statistics. For
the count rates to be expected ($\sim$tens to hundreds of c/s), the
polarisation properties cannot be constrained directly with sub-minute
timescale resolution. Coherent pulsations can be studied simply by
folding the light curve, but folding is not appropriate for stochastic
variability (including QPOs; e.g. \citealt{Ingram2015}). Here, we
present a simple and robust general method for detecting rapid
variability in X-ray polarisation properties, circumventing the
technical challenges associated with stochastic variability. We focus
our analysis mainly on QPOs, but the method can also be used for broad
band variability. In Section \ref{sec:method}, we present our
method. In Section \ref{sec:sims}, we run simulations to determine the
expected signal to noise of the QPOs in polarisation degree and angle
predicted by the precession model. In Section \ref{sec:errors}, we
analyse the improvement in signal to noise that can be achieved by
cross correlating the polarimeter signal with a reference light curve
collected with a large area X-ray detector, before summarising our
results in Section \ref{sec:conc}.

\section{The Method}
\label{sec:method}

In this Section, we first outline why a special method is required for
fast X-ray polarimetry-timing before presenting our method.


\begin{figure*}
	\includegraphics[angle=0,width=\columnwidth]{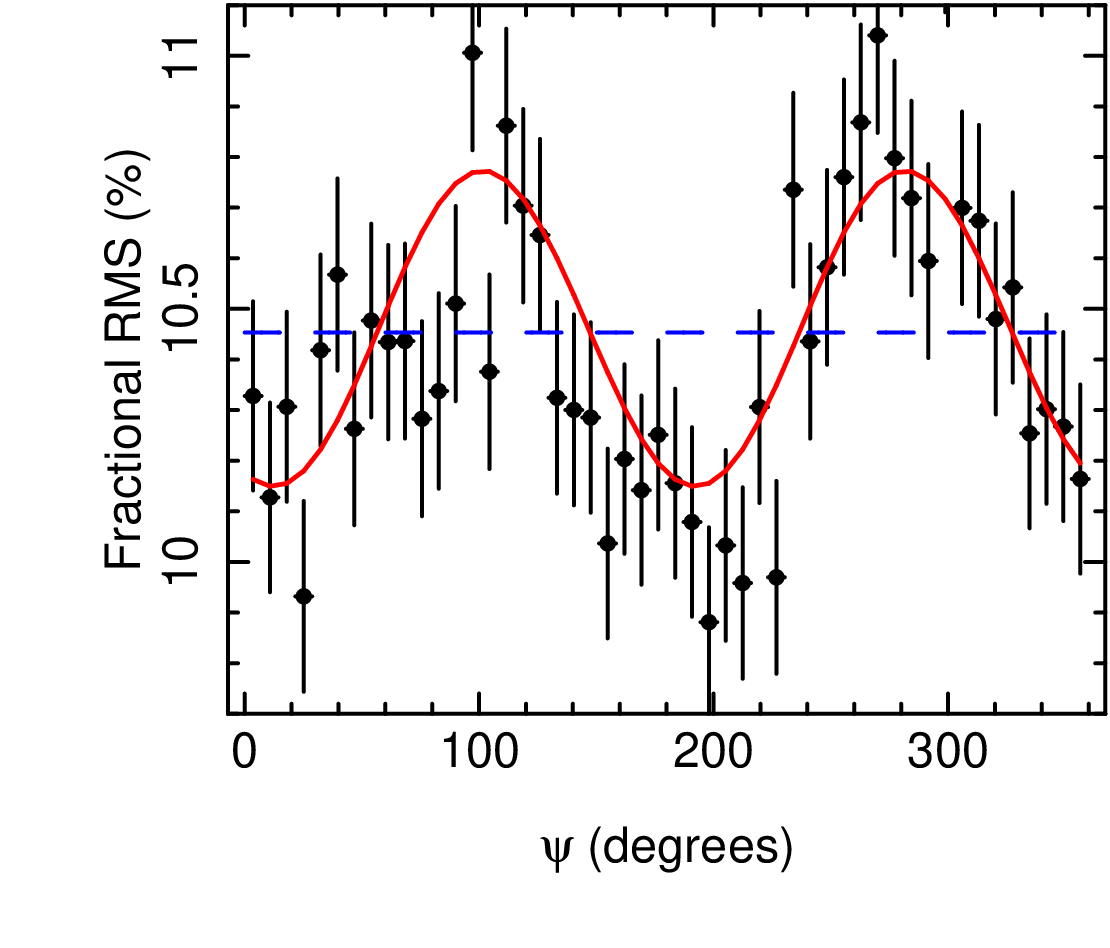} ~~~~
	\includegraphics[angle=0,width=\columnwidth]{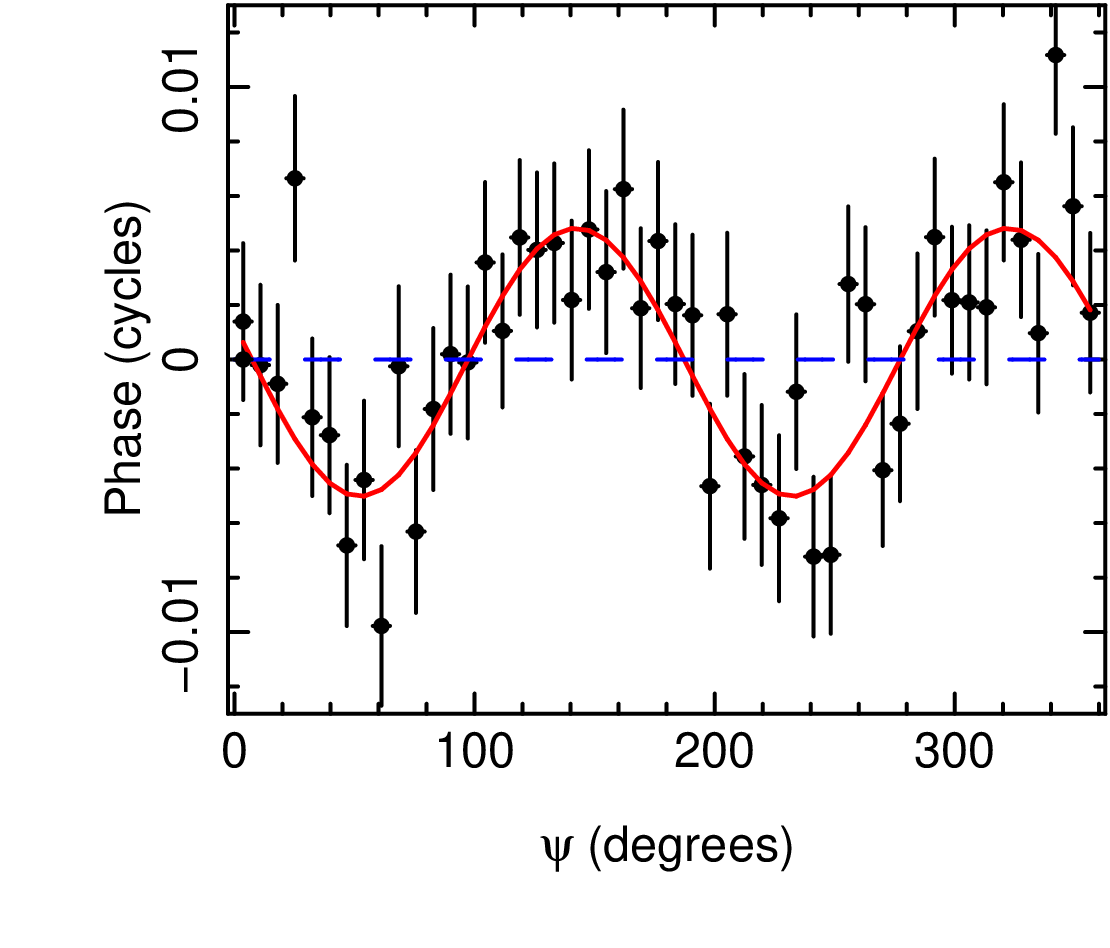}
\vspace{-5mm}
 \caption{Fractional variability amplitude (left) and phase lag
  (right) as a function of modulation angle, $\psi$, for the QPO
  fundamental frequency. The red solid lines assume the oscillations
   in flux, polarisation degree and polarisation angle calculated for the
   high inclination ($i=70^\circ$) model shown in Fig 6 of
  \citet{Ingram2015a} (solid lines therein). The blue dashed lines result
   from assuming the same QPO in the flux, but constant polarisation
   degree and angle. Sinusoidal modulations such as those depicted by
  the red lines therefore provide a robust diagnostic of variability
   in polarisation properties. The black points are a simulation of a
   200 ks \ixpe~exposure, assuming the red lines as the input model.}
 \label{fig:i70}
\end{figure*}

\begin{figure*}
	\includegraphics[angle=0,width=\columnwidth]{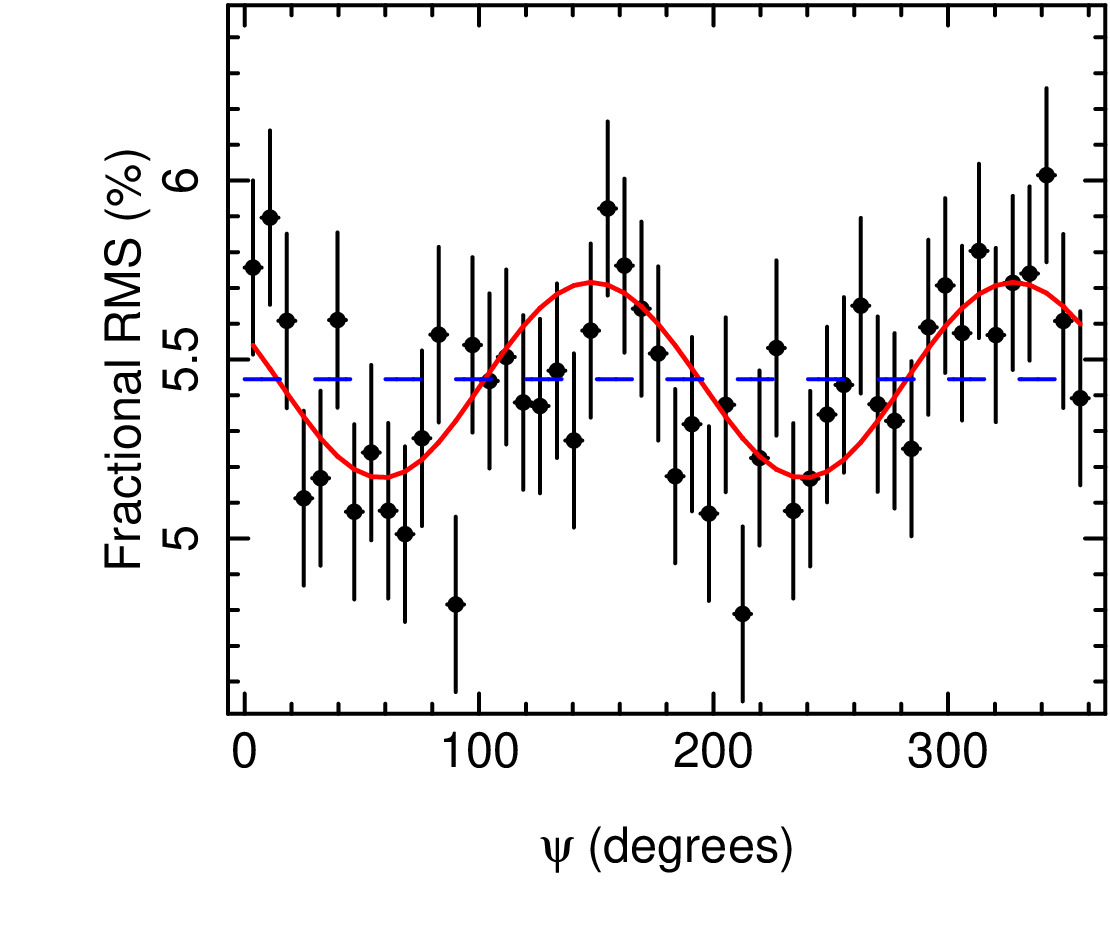} ~~~~
	\includegraphics[angle=0,width=\columnwidth]{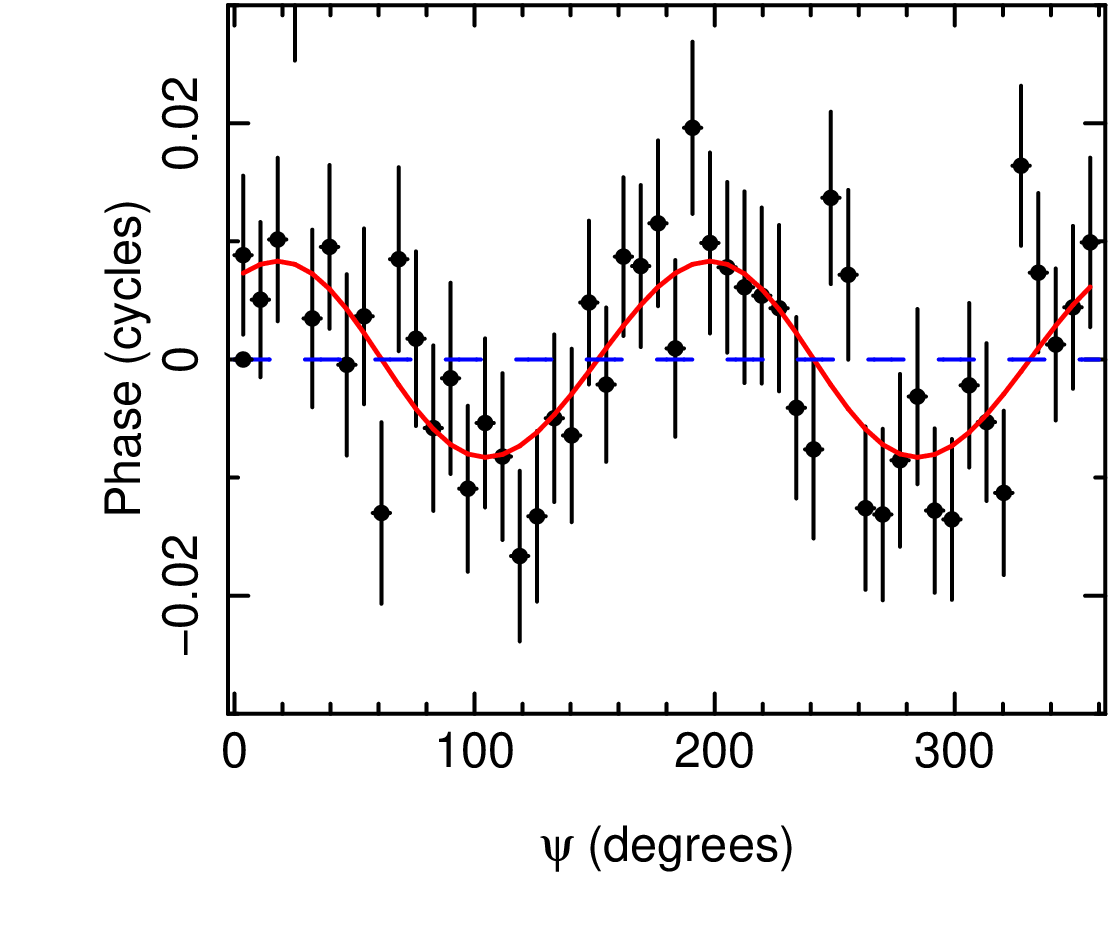}
\vspace{-5mm}
 \caption{The same as for Fig. \ref{fig:i70}, except we now use as
   input the low inclination ($i=30^\circ$) model shown in Fig 8 of
   \citet{Ingram2015a} (solid lines therein).}
 \label{fig:i30}
\end{figure*}

\subsection{The Problem}
\label{sec:problem}

We wish to detect fast ($<10$ s) variability of the
polarisation degree $p_0$ and angle $\psi_0$. This is not trivial,
since we measure these parameters in a statistical
sense by detecting many photons. For each photon, a \textit{modulation
  angle}, $\psi$, is measured, which is effectively an estimate for
the instantaneous polarisation angle of the population of photons. For a
photoelectric effect polarimeter such as the gas pixel detectors
(GPDs) used on \ixpe, this measurement of $\psi$ is obtained from the
orientation of the electron tracks on the 
detector. Similar GPD detectors are also planned to be onboard the
proposed missions \textit{The X-ray Imaging Polarimetry Explorer}
(\xipe; \citealt{Soffitta2016}) and \textit{the Extended X-ray Timing
  and Polarimetry mission} (\extp; \citealt{Zhang2016}). For a Thomson
scattering polarimeter such as the balloon experiment
\textit{X-Calibur} (\citealt{Guo2013}), the measurement is
instead obtained from the position on the detector where the photon
lands. After enough photons have been collected, the polarisation
properties can be measured from a histogram of photon counts versus
modulation angle $\psi$. Specifically, the detected counts as a
function of $\psi$ will be proportional to the modulation function
\begin{equation}
f(\psi|\psi_0,p_0,\mu) = \frac{1}{2\pi} \bigg\{ 1 + \mu ~ p_0
\cos[2(\psi_0-\psi)]  \bigg\},
\label{eqn:fpsi}
\end{equation}
where $\mu$ is the modulation factor of the polarimeter, defined by
the detector's response to a $100\%$ polarised signal. Throughout this
paper, we assume a modulation factor of $\mu=0.3$, which is expected
for \ixpe~(\citealt{Weisskopf2016})\footnote{Also see the
  \ixpe~\textit{WebPIMMS}:
  \url{https://wwwastro.msfc.nasa.gov/ixpe/for_scientists/pimms/}
}. The polarisation degree $p_0$ can therefore be measured from the
amplitude of the modulation  function, and the polarisation angle
$\psi_0$ can be measured from the location of the peak of the
modulation function. Alternatively Stokes parameters can be used,
which is essentially equivalent to measuring the shape of the
modulation function. Note that the modulation function is cyclical on
the interval $\psi=0$ to $\psi=180^\circ$, and therefore $\psi_0$ is
only usefully defined on the interval $0$ to $180^\circ$ (or any
interval spanning $180^\circ$). This is because a wave with a
polarization angle of $\psi_0$ is indistinguishable from a wave with a
polarization angle of $\psi_0+180^\circ$. For example, rotating a
vertically polarised wave by $180^\circ$ leaves another vertically
polarised wave. The modulation angle $\psi$, however, is defined on
the interval $0$ to $360^\circ$ (or any interval spanning
$360^\circ$). For photoelectric effect  polarimeters, this is because
the electron track caused by an incoming photon \textit{does} have a
direction since the starting point of the track can be determined, and
$\psi$ is the angle between the electron track and the projection of
north on the sky. For Thompson scattering polarimeters, $\psi$ also
spans the full $360^\circ$ interval, since it is determined from the
position on the detector where the photon is detected.

The simplest way to measure variability in $p_0$ and $\psi_0$ is of
course to measure both properties directly for many time
intervals. However, detection of polarisation requires a lot of
photons. The minimum detectable polarisation (MDP) is the minimum
polarisation degree that can be detected with statistical confidence
$\mathcal{L}$. This is given by (\citealt{Weisskopf2010})
\begin{equation}
{\rm MDP} = \frac{-\ln(1-\mathcal{L})}{\mu \langle s \rangle }
\sqrt{\frac{\langle s \rangle + \langle b \rangle } {T} },
\end{equation}
where $\langle s \rangle$ and $\langle b \rangle$ represent mean
source and background count rate respectively, and $T$ is the exposure
time. Therefore, for a source with a mean polarisation degree of
$\langle p_0 \rangle = 5\%$, a count rate of $\langle s \rangle =100$
c/s and a negligible background, achieving a statistical confidence of
$\mathcal{L}=99\%$ requires a $T\approx 15$ minute
exposure. For a higher intrinsic polarisation of $\langle p_0 \rangle
= 10\%$, this is still $T \approx 4$ minutes. Therefore, with the
expected count rates and reasonable assumptions about the polarisation
degree, it is not possible to probe sub-minute timescales by directly
calculating time series of $p_0$ and $\psi_0$. To probe faster
timescales, we need a statistical method. For broad band variability
and even QPOs, phase-folding is not a viable method. This is because
the phase of the oscillation does not evolve with time in a
predictable manner.

\subsection{The Solution}
\label{sec:solution}

We can instead consider the variability patterns that will be created
in the modulation function from variability in the count rate,
polarisation angle and polarisation degree. That is, we can make light
curves selected by the modulation angle $\psi$ of each incoming
photon, such that the count rate in the $i^{\rm th}$ $\psi$ bin at
time $t$ is
\begin{equation}
s(\psi_i,t) = s(t) f(\psi_i|\psi_0(t),p_0(t),\mu) \Delta\psi_i,
\label{eqn:tmod}
\end{equation}
where $\Delta\psi_i$ is the width of the $\psi$ bin, $s(t)$ is the
total polarimeter count rate, and the modulation function, $f$, is
given by Equation (\ref{eqn:fpsi}). From Equations (\ref{eqn:tmod})
and (\ref{eqn:fpsi}), it is clear that, if only the total count rate
is varying, and $p_0$ and $\psi_0$ are constant in time, then the
light curves $s(\psi_i,t)$ selected for each $\psi$ bin will all have
the same fractional rms as one another, and will all vary in phase
with one another. This is because the shape of the modulation function
does not vary if $p_0$ and $\psi_0$ remain constant. In contrast, if
only $p_0$ is varying  (with $s$ and
$\psi_0$ now constant), this will cause a stretching and squeezing of
the modulation function as $p_0$ respectively increases and
decreases. This will lead to a peak in fractional rms at $\psi=\psi_0$
and a minimum at $\psi=\psi_0+90^\circ$, with all the light curves
varying in phase with one another as in the previous example. Finally,
if we imagine only $\psi_0$ is varying (with $s$ and $p_0$ constant),
the resulting rocking of the distribution peak will lead to light curves
for $\psi_i > \langle \psi_0 \rangle$ varying in anti-phase with light
curves for $\psi_i < \langle \psi_0 \rangle$. For a more general (and
realistic) situation, with $s$, $p_0$ and $\psi_0$ all varying, an
intuition is harder to form and calculations are required. However, it
is  possible to appreciate that variability of polarisation properties
is encoded in the $\psi$ dependent variability properties of the
signal, which can be probed using standard cross-spectral techniques
developed for the purposes of spectral-timing.

We can therefore define a reference time series, $r(t)$, that is highly
correlated with all the $s(\psi_i,t)$ light curves and define a set of
cross-spectra
\begin{equation}
C(\psi_i,\nu,\Delta) = \langle S(\psi_i,\nu) R^*(\nu) \rangle,
\label{eqn:cross}
\end{equation}
where an uppercase letter represents the Fourier transform (FT) of the
corresponding lowercase letter and a star denotes a complex
conjugate. The angle brackets denote averaging, which is over an
ensemble of different realisations (i.e. the light curves are split
into many segments) and also over the Fourier frequency range
$\nu-\Delta/2$ to $\nu+\Delta/2$ (\citealt{vanderKlis1987}). We see
that the only difference with more familiar spectral-timing analyses,
is that we are selecting light curves based on modulation angle rather
than energy. Everything else is, in principle, entirely equivalent. The
reference light curve may be provided by a second detector on the same
satellite as the polarimeter, such as the $\sim 3 {\rm m}^2$
\textit{Large Area Detector} (LAD) of \extp. Alternatively, it could
be provided by a simultaneous pointing from another observatory, such
as \astrosat~(\citealt{Singh2014}) which is likely to still be in
operation during the \ixpe~mission lifetime, or \textit{The
  Spectroscopic Time-Resolving   Observatory for Broadband   Energy
  X-rays} (\textit{STROBE-X}; \citealt{WilsonHodge2017}) which is
proposed to include an $\sim 8 {\rm m}^2$ version of the LAD. In the
absence of another instrument, the total polarimeter count rate
(i.e. summed over all $\psi$) could be used. It is convenient if the
reference time series is statistically independent from the other
light curves. This property is automatically satisfied by the use of a
second instrument\footnote{Although in practice electronic issues, such
  as very large events that are picked up by more than one detector, can
  lead to statistical independence of detectors being lost.}, and can
be ensured by using, for example, the total
polarimeter count rate minus the currently considered $\psi$ bin in
the absence of a second instrument\footnote{Alternatively, one could
  use e.g. even energy 
  channels for the reference time series and odd energy channels for
  the other light curves, or simply deal with the mathematics of not
  having statistical independence.} (in direct
analogy to spectral-timing techniques;
e.g. \citealt{Uttley2014}). From the cross-spectrum for a given
Fourier frequency range, we can calculate the fractional rms as a
function of $\psi$, and also the phase lag as a function of $\psi$
with respect to the reference time series (See Appendix \ref{sec:bbn}
for more details).

The red solid lines in Fig \ref{fig:i70} and \ref{fig:i30} show the
fractional rms and phase lags as a function of $\psi$ calculated by
inputing the (LF) QPOs in polarimeter count rate, polarisation degree and
angle predicted by \cite{Ingram2015a}. In that paper, the authors ray
trace radiation from a precessing torus to a distant observer using
the Kerr metric, and calculate the resulting polarisation properties
as a function of precession phase. Two parameter combinations are
featured, referred to here as the high inclination model ($i=70^\circ$,
$\Phi=110^\circ$, $\beta=10^\circ$; see Fig 6 in
\citealt{Ingram2015a}) and the low inclination model ($i=30^\circ$,
$\Phi=180^\circ$, $\beta=10^\circ$; see Fig 8 in
\citealt{Ingram2015a}). Note that \citet{Ingram2015a} used the symbol
$\chi$ for the polarisation angle, whereas here we use the symbol $\psi_0$
- reserving $\chi$ for the $\chi^2$ fit statistic. Figs \ref{fig:i70}
and \ref{fig:i30} correspond to the high and low inclination model
respectively. We consider the rms and phase lags at the QPO
fundamental frequency, and we also take onto account the broad band
noise that is observed coincident with Type C QPOs. The details of our
calculation are presented in Appendix \ref{sec:bbn}. The figures show
approximately sinusoidal modulations in both the amplitude and phase
resulting from $p_0$ and $\psi_0$ varying with QPO phase. The blue
dashed lines show an alternative, null-hypothesis, model, in which
only the count rate varies and $p_0$ and $\psi_0$ stay constant with
QPO phase. As expected, we see no modulations in either the amplitude
or phase for this null-hypothesis model. We therefore have a simple
and statistically robust way to detect variability in the polarisation
properties: simply by looking for these $\sim$sinusoidal modulations
in the rms and phase lags as a function of $\psi$. We note that just
detecting these modulations does not automatically tell us about
whether it is $p_0$, $\psi_0$ or both varying. This requires a more
detailed analysis (see Section \ref{sec:polang}). There is one
exception however. If $\psi_0$ is constant and $p_0$ varies in phase
with the total count rate, then there will be a sinusoidal modulation
in the amplitude but not in the phase lags. We finally note that the
rms and phase are both cyclical on the interval $0$ to
$180^\circ$, which is because $\psi_0$ is only defined on an interval
of $180^\circ$ (see Section \ref{sec:problem}).

\section{Simulations}
\label{sec:sims}

In this section, we present methods to detect $p_0$ and $\psi_0$
oscillations in noisy data, with null-hypothesis significance
testing. Throughout, we consider a $200$ ks exposure of a bright
source (absorbed power-law spectrum with index $\Gamma=2$,
normalisation $=3$ photons/s/cm$^{2}$/keV, hydrogen column density
$n_h=10^{22}$ cm$^{-1}$) with negligible background, comparable to
e.g. GX 339-4 in a bright hard state or intermediate state. We first
describe our simulation method, focusing on \ixpe. We then introduce a
simple null-hypothesis test, which compares a sinusoidal model for the
rms and phase lag modulations as a function of $\psi$ to a
null-hypothesis model with constant rms and phase lag. This determines
the statistical confidence with which we prefer a model with variable
polarisation properties over a null-hypothesis model with constant
polarisation properties. We then specifically consider how to
constrain an oscillation in polarisation angle, which is the most
interesting quantity, offering a `smoking gun' detection of
precession.

The null-hypothesis tests presented here additionally allow us to
properly explore trade-offs when deciding whether to target high or
low inclination sources to search for an oscillation in polarisation
angle. In the precession model (and in the
observational data: \citealt{Schnittman2006,Motta2015,Heil2015}), the
oscillation in the flux has a larger amplitude for higher
inclinations (i.e. systems viewed more edge-on), since there is more
variability in solid angle and Doppler boosting over each precession
cycle. The mean polarisation degree is also expected to increase with
inclination angle (\citealt{Chandrasekhar1960,Sunyaev1985}). However,
the oscillation in polarisation angle has a greater amplitude for the
low inclination model, since a precessing vector traces out a cone
when viewed from the side and a full circle when viewed from the
top. We therefore consider both the high and low inclination model in
this Section.

\subsection{Simulation setup}

We present details of our simulations in Appendix
\ref{sec:simdet}. Here we summarise the general scheme. We calculate
our model for the fractional rms and phase lag corresponding to each
QPO harmonic as a function of $\psi$ as described in Appendix
\ref{sec:bbn}. We generate synthetic data by calculating $1~\sigma$
errors on the model (the expression for which we present and discuss
in Appendix \ref{sec:simdet}) and selecting Gaussian random
variables. With real data, we would measure the rms and phase lags
by calculating the cross-spectrum for many segments, each of length
$T_{\rm seg}$ seconds, and averaging. We would also average over all
the frequency bins lying in the frequency range $\Delta$ Hz. Since
there are a total of $T/T_{\rm seg}$ segments (where $T$ is the total
exposure time), and the frequency resolution is $d\nu=1/T_{\rm seg}$
(\citealt{vanderKlis1989}), the averaging is over $T \Delta$
realisations of the cross-spectrum. Therefore $T$ and $\Delta$ are
important parameters for calculating the error on the rms and phase
lags. For a QPO, it is appropriate to average over the frequency range
$\nu_k - \Delta_k/2$ to $\nu_k + \Delta_k/2$, where $\nu_k$ and
$\Delta_k$ are respectively the centroid frequency and full width at
half maximum (FWHM) of the $k^{\rm th}$ QPO harmonic. These two
parameters are related by the quality factor, $Q=\nu_k/\Delta_k$,
which is generally observed to be $Q\sim 8-10$ for most Type C LF QPOs
(note that the quality factor is generally equal for all detected
harmonics). The remaining inputs required to calculate the errors are
the mean polarimeter and reference time series count rates, $\langle s
\rangle$ and $\langle r \rangle$. These parameters determine the
Poisson noise level.

The black points in Fig \ref{fig:i70} and \ref{fig:i30} show our
simulation results for a $T=200$ ks simulated exposure with only \ixpe,
using the red solid lines as the input model. The polarimeter
mean count rate is $\langle s \rangle =100$ c/s, calculated assuming
the spectral parameters defined at the start of this Section and
folding around the \ixpe~response matrix. We set the reference time
series count rate to $\langle r \rangle = \langle s \rangle$, assuming
that we can use the total \ixpe~count rate as the reference time
series. We set $\Delta =0.2$ Hz, which is appropriate for a QPO with
centroid frequency $\nu=1.6$ Hz and quality factor $Q = 8$. For the
high inclination model (Fig \ref{fig:i70}), the predicted modulations
in both fractional rms and phase lag are clearly visible. For the low
inclination model (Fig \ref{fig:i30}), the modulations are less clear
in the rms, but can be seen in the larger amplitude phase lag
modulations (resulting from the larger amplitude oscillations in
polarisation angle in the low inclination model). We only plot the
results for the QPO fundamental here. The modulations are not
detectable in the synthetic data for higher harmonics.

\subsection{Null-hypothesis testing}
\label{sec:null}

We now formally test the confidence with which we can rule out  a
null-hypothesis of constant $p_0$ and $\psi_0$ for the synthetic data
points in Fig \ref{fig:i70} and \ref{fig:i30}. For this
null-hypothesis, the fractional rms and phase would not depend on
$\psi$. Alternatively, for a model in which $p_0$ and/or $\psi_0$ are
varying, the fractional rms and phase would have an approximately
sinusoidal dependence on $\psi$. We can therefore fit two models to
the data and, since one model is a `nested' version of the other,
compare the goodness of fit using an F-test. For the null-hypothesis,
we simply calculate the error weighted mean fractional amplitude and
phase lag from the data (this is identical finding a best-fitting
constant by minimising $\chi^2$). For the `full' model, we fit a
sinusoid function to both fractional rms and phase lag
\begin{equation}
y(\psi) = A + B \cos[ 2 ( \psi - C ) ],
\end{equation}
where $A$, $B$ and $C$ are free parameters in each of the two fits
(i.e. one fit to the rms and the other to the lags).

For the high
inclination simulation shown in Fig \ref{fig:i70}, the best fitting
null-hypothesis model has a reduced $\chi^2$ of
$\chi^2_\nu=234.6/98$. This is calculated for both rms and phase over
a total of 100 data points (50 rms points and 50 lag points), with
only two free parameters (the mean rms and the mean phase lag),
resulting in 98 degrees of freedom. The sinusoidal model has a much
better reduced $\chi^2$ of $98.6/94$. Here, there are the same number
of data points but now there are 6 free parameters ($A$, $B$ and $C$
for rms and phase lag), giving 94 degrees of freedom. An F-test
returns an F statistic of $F=32.4$, which corresponds to a null-hypothesis
p-value far lower than the threshold for $5 \sigma$ confidence
($p=6\times 10^{-17}$). For the low inclination simulation shown in
Fig \ref{fig:i30}, the reduced $\chi^2$ values for null-hypothesis and
sinusoidal models are instead $\chi^2_\nu=158.2/98$ and
$\chi^2_\nu=85.8/94$ respectively. This again gives a p-value
corresponding to $>5\sigma$ confidence ($F=19.8$, $p=7\times
10^{-12}$). As a final check, we also compare the input model with
the data, to get $\chi^2_\nu=104.9/100$ and $\chi^2_\nu=90.0/100$ for the high and
low inclination models respectively, indicating good fits.

We note that $\chi^2$ statistics are only appropriate for the case of
Gaussian errors. This is clearly the case for our simulation, since we
select Gaussian random variables. In practice however, the
cross-spectra must be averaged over a suitably large number of
realisations for the Gaussian limit to be reached (i.e. the central
limit theorem). This requires $T\Delta \gtrsim 400$
(\citealt{Vaughan2003}), which is comfortably the case for our chosen
parameters ($T\Delta = 2\times10^4$).

Thus, we expect to be able to detect the
oscillations predicted by \cite{Ingram2015a}, even for low inclination
sources. However, there is of course some uncertainty over what we
expect theoretically for the $p_0$ and $\psi_0$ oscillations. It is
therefore worth exploring parameter space with our new simple
hypothesis testing tool. The most robust prediction of the precession
model is the $\psi_0$ oscillation, since this largely depends on
geometry alone. The modulation in $p_0$ is much more uncertain, since
it depends on the angular dependence of emergent radiation, which in
turn depends on the details of the Comptonisation process that drives
the hard X-ray radiation. A key uncertainty is the average
polarisation degree, $\langle p_0 \rangle$. In particular, the
calculations of \cite{Sunyaev1985} used by \cite{Ingram2015a} may
over-estimate $\langle p_0 \rangle$, since they only consider photons
that have had `many scatterings'. More detailed calculations have
since predicted lower polarisation from Comptonisation
(e.g. \citealt{Schnittman2010}), particularly in the $2-8$ keV range
that the GPD detectors are sensitive to. This is largely because
photons with energies of $2-8$ keV often do not fulfil the
aforementioned criterion of having been scattered many times. It is
worth mentioning that polarimeters sensitive to harder X-rays are
therefore desirable, since more $\sim 10-50$ keV photons will have
undergone many scatterings and are therefore expected to be more
highly polarised (also the LF QPO fractional amplitude is often observed
to increase with energy). Thomson scattering polarimeters such as the
\textit{Polarization Spectroscopic Telescope Array} (\textit{PolSTAR}:
\citealt{Henric2016}), the proposed satellite version of
\textit{X-Callibur}, are therefore promising prospects for the future.

\begin{figure*}
	\includegraphics[angle=0,width=\columnwidth]{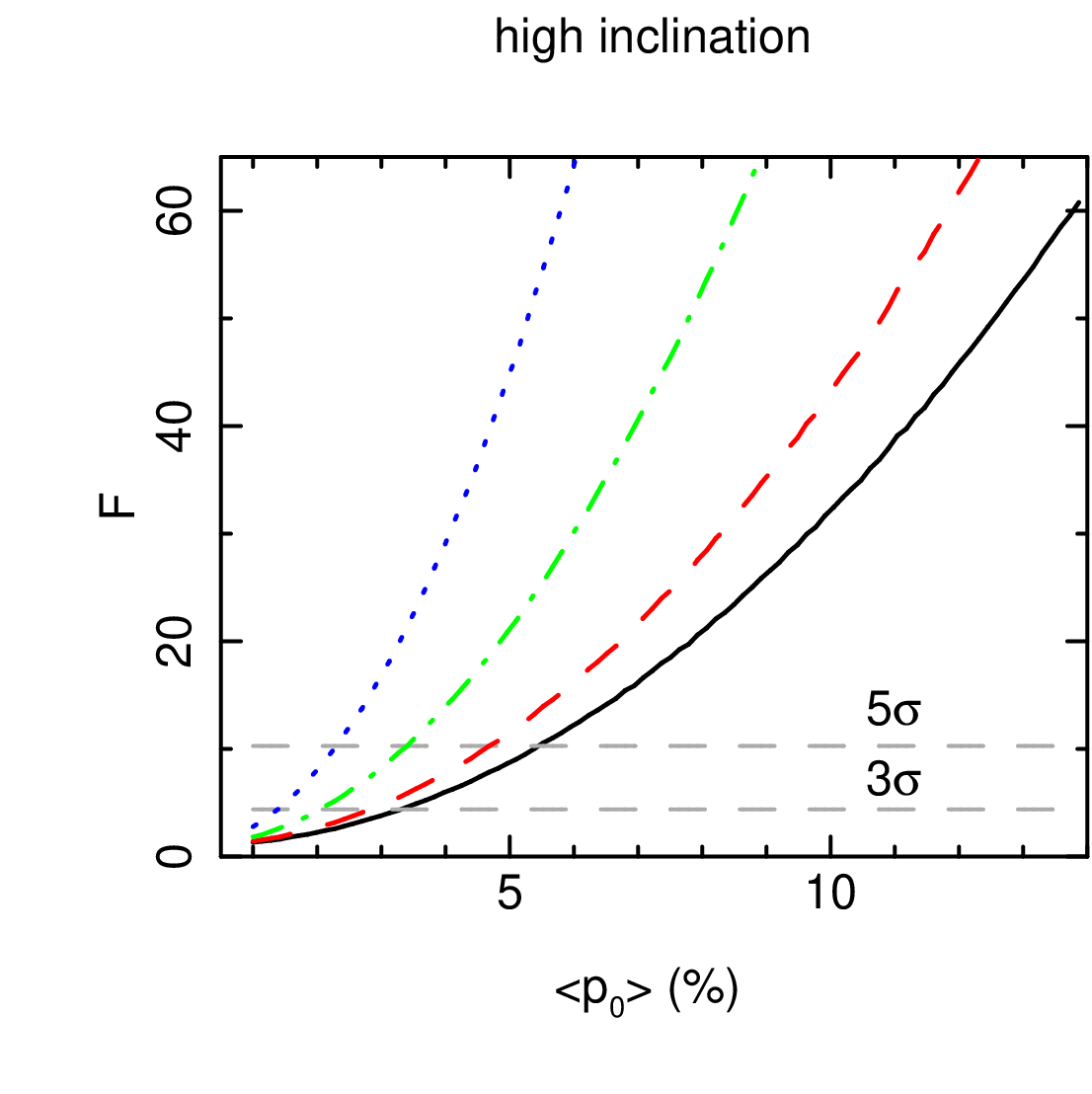} ~~~~
	\includegraphics[angle=0,width=\columnwidth]{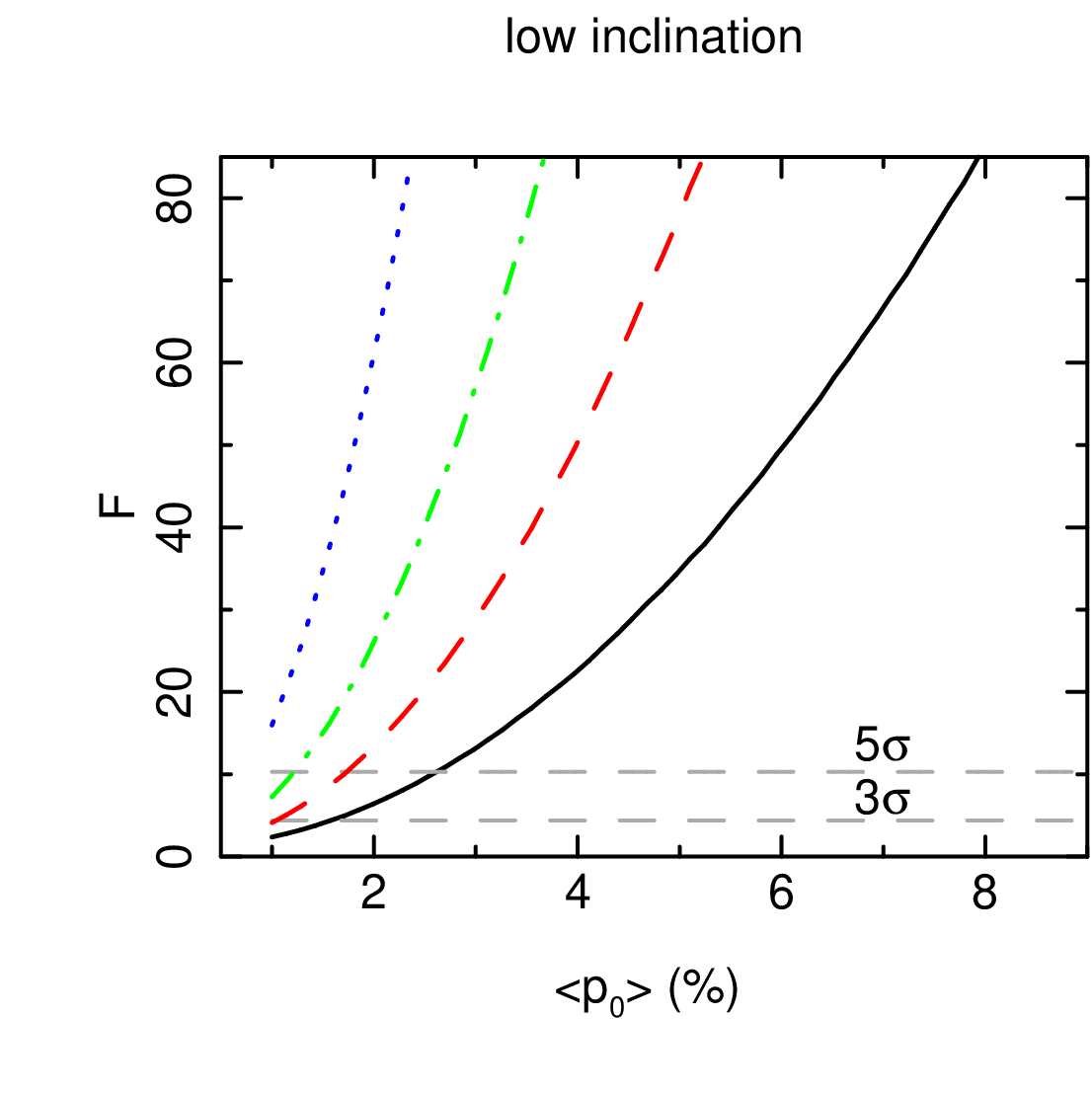}
\vspace{-7mm}
 \caption{The curves show the $F$ statistic calculated by
   comparing the best-fitting sinusoid model with a null-hypothesis
   corresponding to constant polarisation properties (see text
   for further details) as a function of mean polarisation
   degree. The left and right plots are for the simulations
   corresponding to the high and low inclination models
   respectively. Each curve represents a different combination of mean
   count rate in the polarimeter $\langle s \rangle$ and mean count rate in
   the reference time series, $\langle r \rangle$. From bottom to top,
   they represent $\langle s \rangle =100$ c/s, $\langle r
   \rangle=100$ c/s (black solid), $\langle s \rangle=100$, $\langle r
   \rangle=5000$ (red dashed), $\langle s \rangle =200$, $\langle r
   \rangle=38,000$ (green dot-dashed) and $\langle s \rangle=500$, $\langle r
   \rangle =38,000$ (blue dotted). These cases are respectively relevant for 
   \ixpe~alone, \ixpe+\astrosat, \extp~baseline and \extp~goal. The
   grey dashed lines show $3 \sigma$ and $5 \sigma$ values. We see that,
   for the high inclination model, a $5 \sigma$ detection is possible
   for $\langle p_0 \rangle \gtrsim 5.5 \%$ with \ixpe~alone, or
   $\langle p_0 \rangle \gtrsim 2.3 \%$ with the \extp~goal
   configuration. Smaller polarisation degrees are required for the
   low inclination model, but we do expect low inclination sources to
   be less polarised.}
 \label{fig:X2}
\end{figure*}

In Fig. \ref{fig:X2}, we therefore explore a range of $\langle p_0
\rangle$ values, again considering a $200$ ks exposure and
$\Delta=0.2$ Hz. We use the same high (left) and (low) inclination
input models as before for $s$, $p_0$ and $\psi_0$ as a function of QPO
phase, except we re-scale the $p_0$ oscillation by the new mean.
The curves show the $F$ statistic resulting from comparing
a null-hypothesis and a sinusoidal model to synthetic data for
different polarimeter and reference time series count rates. In order
to smooth out noise, we average $\chi^2$ values over $10,000$
realisations of synthetic data before doing the F-test. From
bottom to top, the curves represent: only \ixpe~(black solid),
\ixpe~with \astrosat~recording the reference time series (red dashed),
\extp~`requirement' specifications with the reference time series
recorded by the $\sim 3$ m$^2$ LAD (green dot-dashed) \footnote{Here,
  the requirement specifications of the \extp~polarimeter are assumed
  to be the same as the \xipe~specifications. Therefore the green
  dot-dashed line could also represent \xipe~plus the $\sim 3$ m$^2$ LAD.}, and \extp~`goal'
specifications (blue dotted). The grey dashed lines represent $3$ and $5~\sigma$
confidence. As expected, the oscillations in $p_0$  and $\psi_0$ are
harder to detect when the mean polarisation degree is low. Also, for
lower inclination angles, the minimum degree of polarisation required
for a significant detection of polarisation variability is
smaller. This is because the swings in polarisation angle are
predicted to have a larger amplitude for the low inclination
model. However, we \textit{do} expect a lower mean
polarisation degree for lower inclinations, so it is likely still best
to target high inclination sources. We see that sensitivity is
improved by using configurations that increase the mean polarimeter 
count rate $\langle s \rangle$ and the mean reference time series
count rate $\langle r \rangle$. Therefore, for a given polarimeter, we
can increase signal to noise simply by observing simultaneously with
another observatory with greater collecting area than the
polarimeter. We will discuss the relative importance of high reference
and polarimeter count rates in Section \ref{sec:errors}.

\subsection{Measuring an oscillation in polarisation angle}
\label{sec:polang}

\begin{figure*}
	\includegraphics[angle=0,width=\columnwidth]{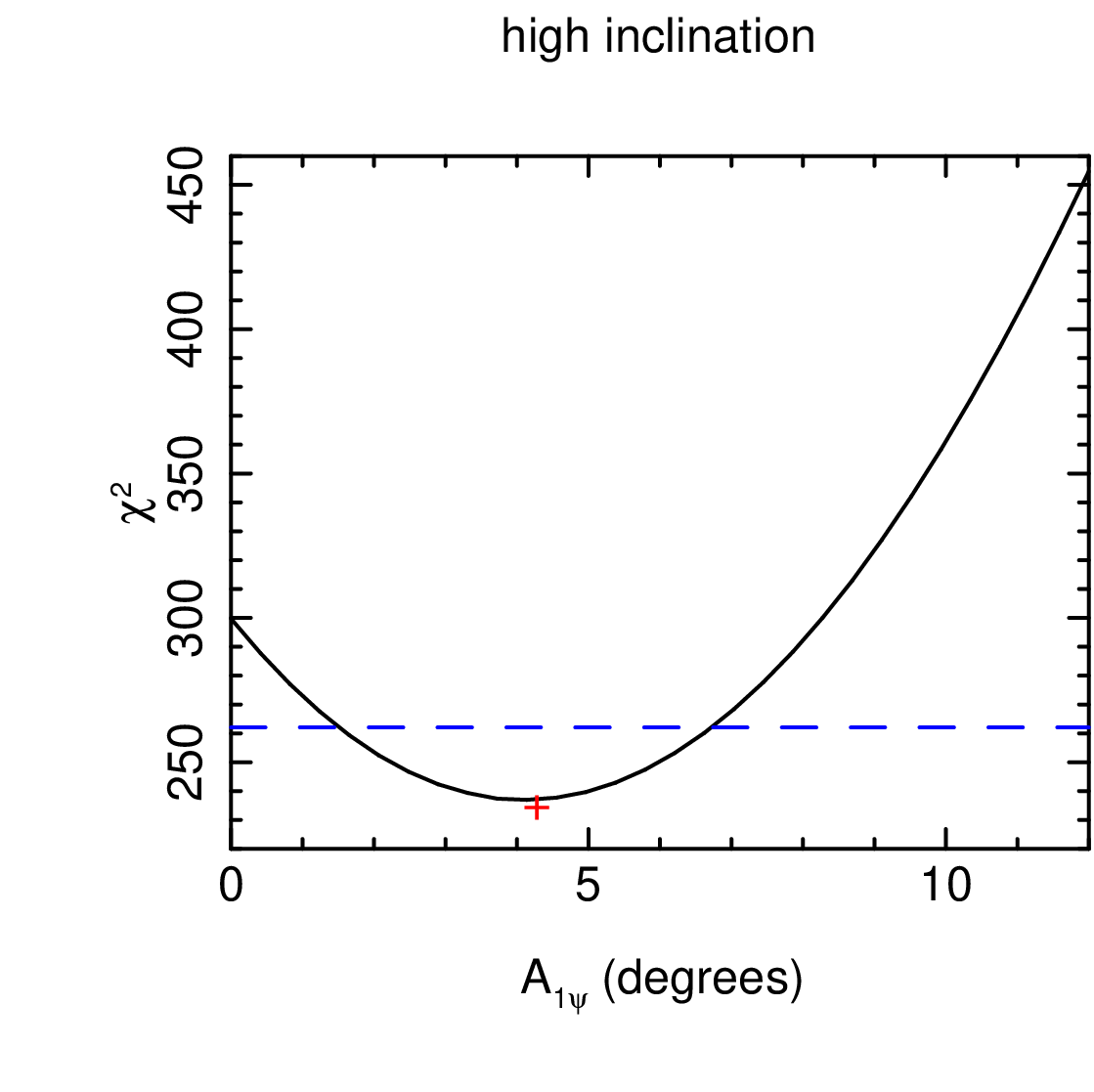} ~~~~
	\includegraphics[angle=0,width=\columnwidth]{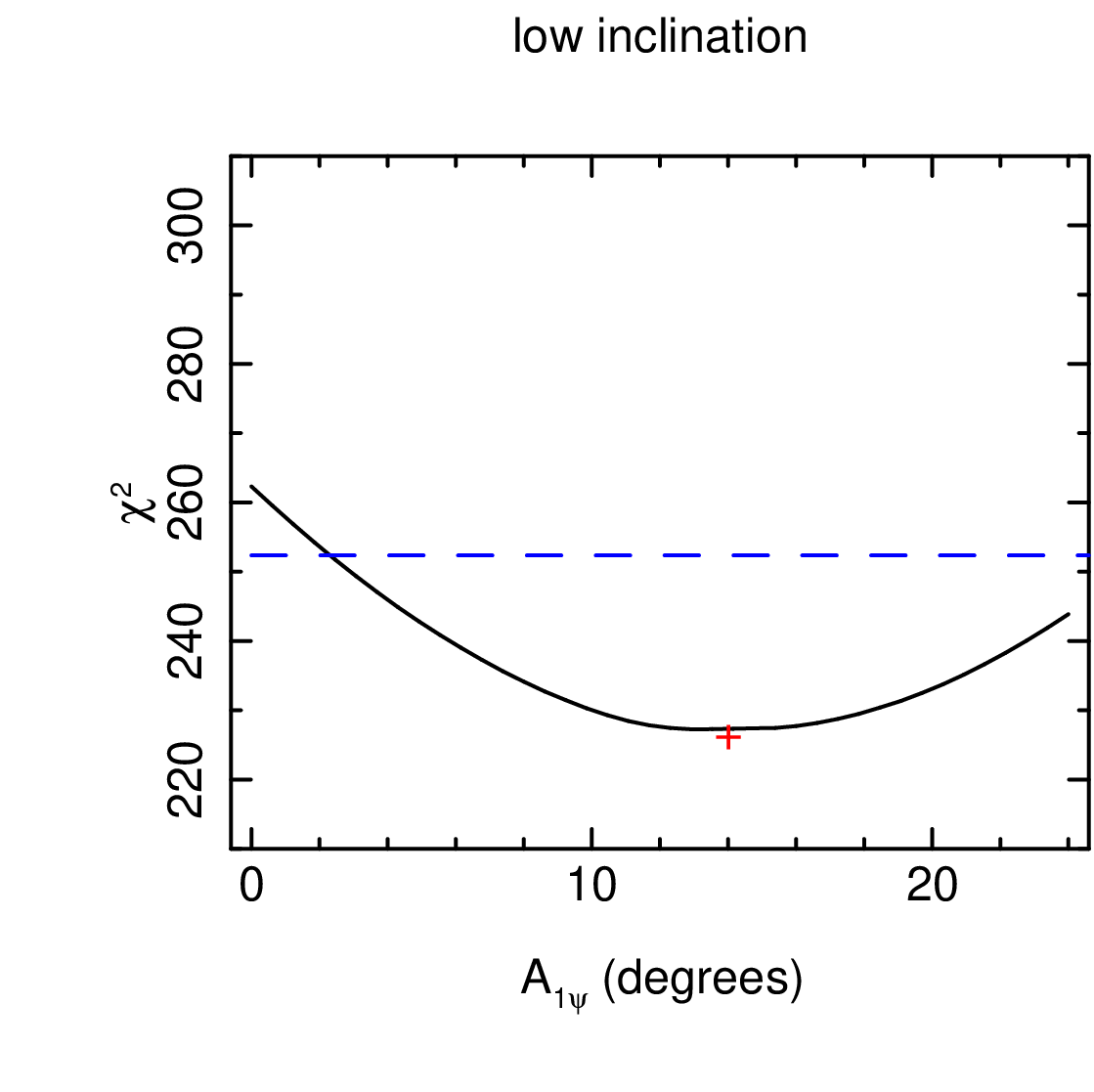}
\vspace{-5mm}
 \caption{Plot of minimum $\chi^2$ against the parameter $A_{1\psi}$
   (black solid line) for fits to synthesised \ixpe~data. A non-zero
   value of this parameter indicates that the polarisation angle is modulated on the
   QPO fundamental. The left and right hand panels correspond to the
   high and low inclination models respectively. The red crosses show
   the input value of $A_{1\psi}$ for each simulation and the blue
   dashed lines show the $5 \sigma$ confidence contour. For both input
   models the null-hypothesis of $A_{1\psi}=0$ can be strongly ruled out.}
 \label{fig:contour}
\end{figure*}

We have shown that it is simple to measure variability in polarisation
properties with an X-ray polarimeter. However, determining if
specifically the polarisation angle is oscillating, and/or if the
polarisation degree is oscillating, and moreover measuring the
amplitude and phase of those oscillations, requires further work. In
this section, we show that a method entirely analogous to the
\cite{Ingram2016} QPO phase-resolving method can be employed to do
just this. We start by representing the oscillations in count rate,
polarisation degree and angle as a simple phenomenological function of
QPO phase, $\omega$. For example, the $\psi_0$ oscillation is given by
\begin{equation}
\psi_0(\omega) = \langle \psi_0 \rangle + A_{1\psi_0}
\sin[\omega - \phi_{1\psi_0}] + A_{2\psi_0}
\sin[2(\omega - \phi_{2\psi_0})],
\end{equation}
and we use equivalent expressions for $p_0(\omega)$ and
$s(\omega)$. This is simply a sum of harmonics, where we only
consider two harmonics (since in most cases only two QPO harmonics
can be detected). Here, $A_{1\psi_0}$ and $A_{2\psi_0}$ are the
amplitudes of respectively the first and second harmonics of the
$\psi_0$ oscillation, and $\phi_{1\psi_0}$ and $\phi_{2\psi_0}$ are
the phases. Including the 3 mean parameters ($\langle s \rangle$,
$\langle p_0 \rangle$ and $\langle \psi_0 \rangle$), 6 amplitude
parameters [i.e. $A_{1\psi_0}$, $A_{2\psi_0}$ and the equivalents for
$p_0(\omega)$ and $s(\omega)$], and 6 phase parameters, there are 15
model parameters altogether. We simultaneously fit this model to the
synthetic data of the mean count rate vs $\psi$, the fractional rms vs
$\psi$ for two QPO harmonics, and the phase lag vs $\psi$ for two QPO
harmonics. The simultaneous fit is therefore performed over a total of
five datasets. We use \textsc{xspec} v 12.9 (\citealt{Arnaud1996}).

We use the same two simulations shown in Figs. \ref{fig:i70} and
\ref{fig:i30}. Although these figures only show the rms and phase for
the QPO fundamental (i.e. first harmonic), we also consider the second
harmonic and the mean count rate in our fit. Even though the synthetic
data for the second harmonic are very noisy, they still constrain our
model since they rule out parameter combinations that predict very
large rms and lag modulations in the second harmonic that are not
present in the synthetic data. Fig. \ref{fig:contour} shows the
minimum $\chi^2$ as a function of the parameter $A_{1\psi_0}$ (black
solid line) for the high (left) and low (right) inclination models. In
both cases, the red cross depicts the exact answer calculated directly
from the input model for the simulation, which is clearly consistent
with the best fit value. The blue dashed line depicts the $5~\sigma$
confidence contour. We see that the QPO in $\psi_0$ is detected with
$> 5 \sigma$ confidence in both cases, with the significance being
greater for the high inclination model. We note that the simulation
includes the effect of broad band noise, but our simple model fitting
in this Section does not. The fact that we recover the input
polarisation angle oscillation accurately gives us confidence that the
method is fairly robust to biases introduced by the broad band noise
signal. Clearly, it is also possible to use the same method presented
here to measure the oscillation in polarisation degree.

%



\begin{figure*}
	\includegraphics[angle=0,width=\columnwidth]{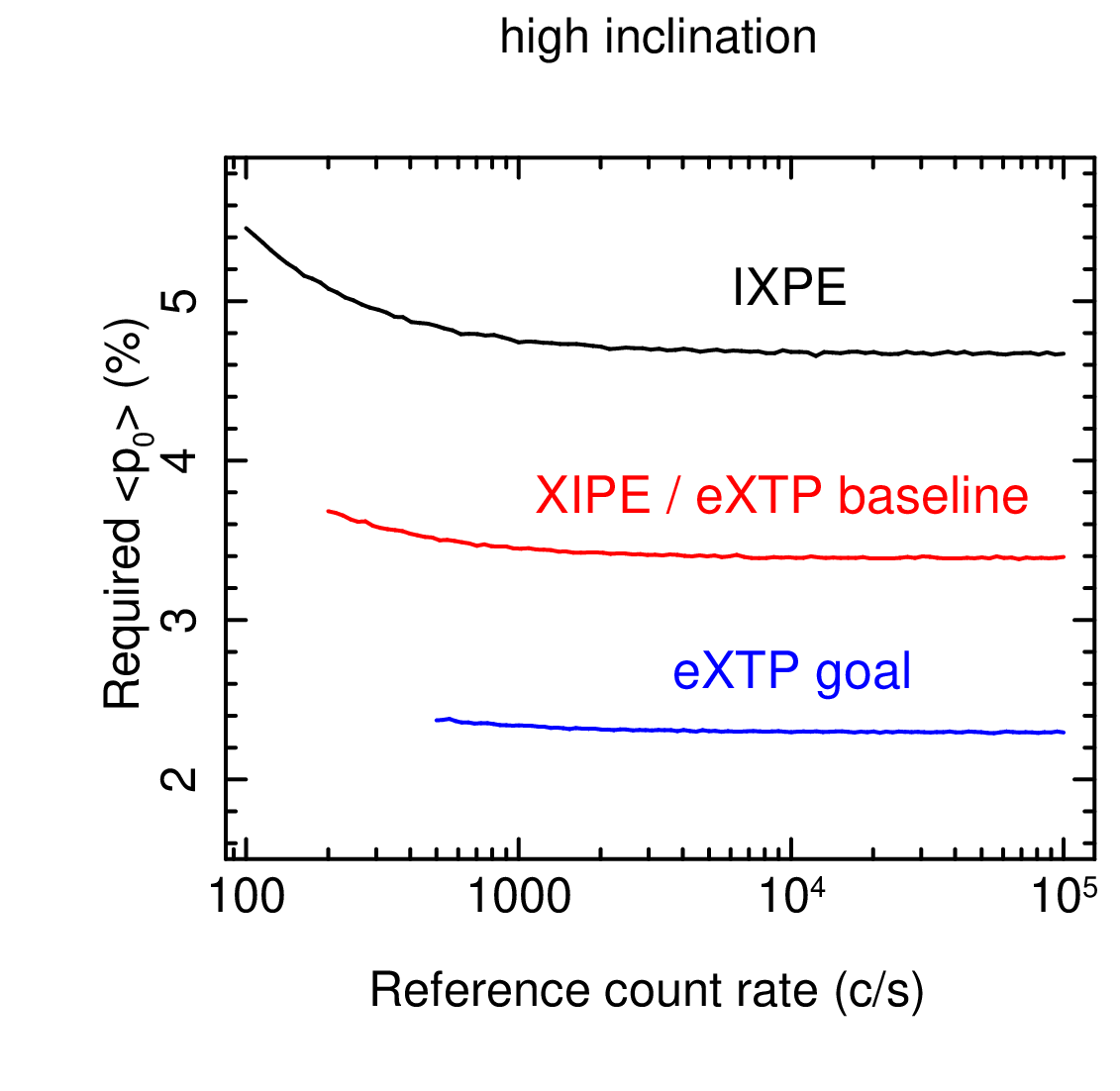} ~~~~
	\includegraphics[angle=0,width=\columnwidth]{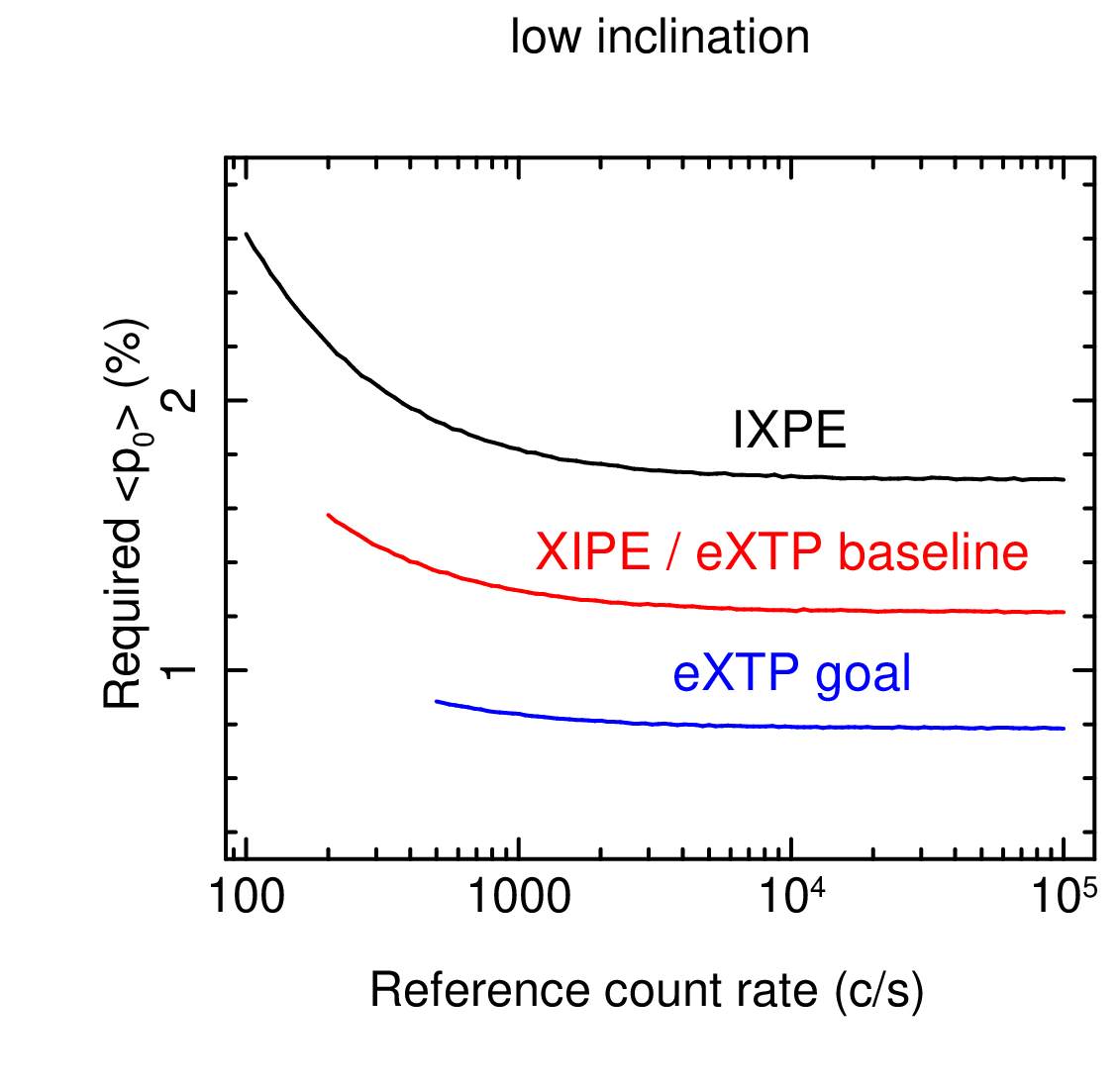}
\vspace{-5mm}
 \caption{Mean polarisation degree required in order to make a
   $5\sigma$ detection of the oscillations in polarisation properties
   predicted by \citet{Ingram2015a} for a high (left) and low (right)
   inclination object, plotted against mean count rate of the reference
   time series. We assume a 200 ks exposure and consider three
   different specifications of polarimeter, assuming mean polarimeter
   count rates of $\langle s \rangle = 100$ c/s (black: \ixpe),
   $\langle s \rangle = 200$ c/s (red: \xipe~/ \extp~baseline) and
   $\langle s \rangle = 500$ c/s (blue: \extp~goal). We see that
   increasing the polarimeter count rate increases sensitivity, as
   does increasing the reference count rate. We can therefore
   increase the sensitivity of a given polarimeter by using a large
   area instrument to collect the reference time series. There is
   however a saturation point (here at $\sim 5000$ c/s) beyond which
   increasing the reference count rate provides no further advantage.}
 \label{fig:pdet}
\end{figure*}

\begin{figure*}
	\includegraphics[angle=0,width=\columnwidth]{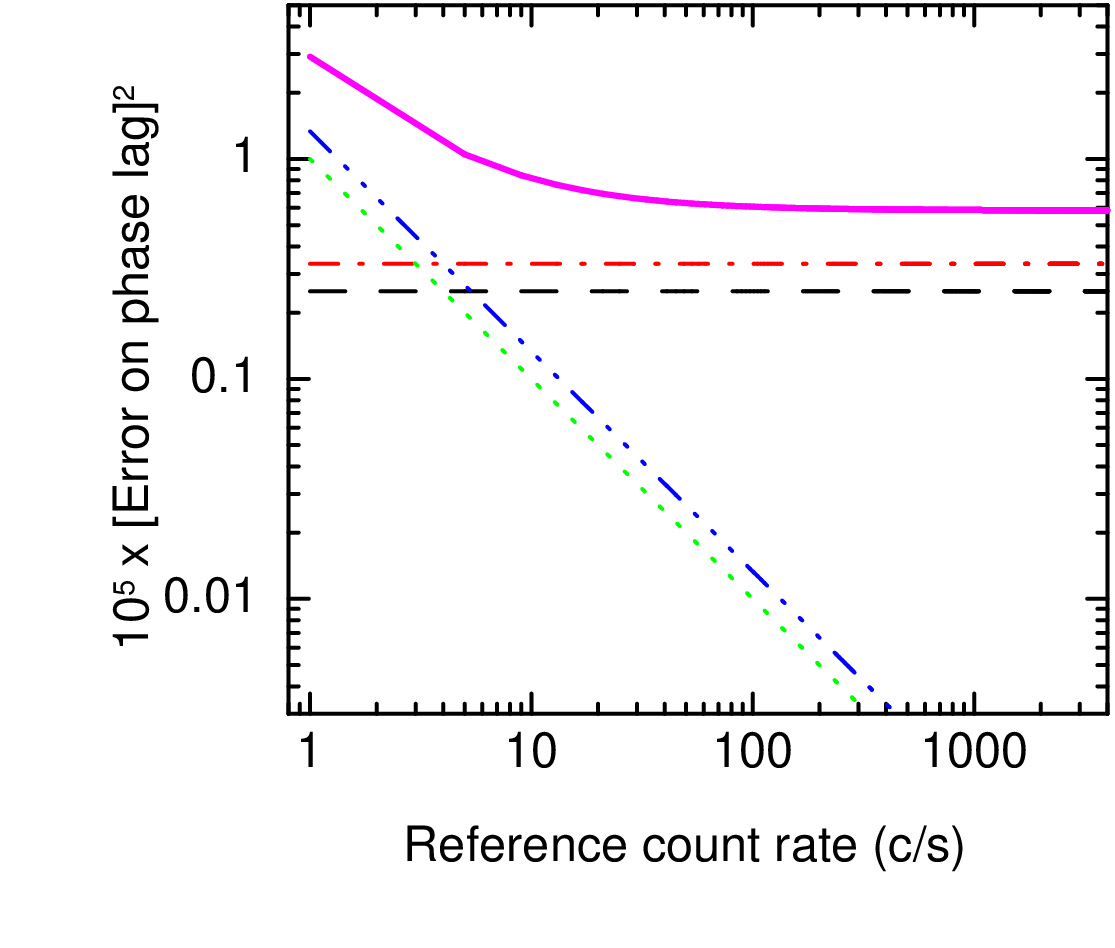} ~~~~~
	\includegraphics[angle=0,width=\columnwidth]{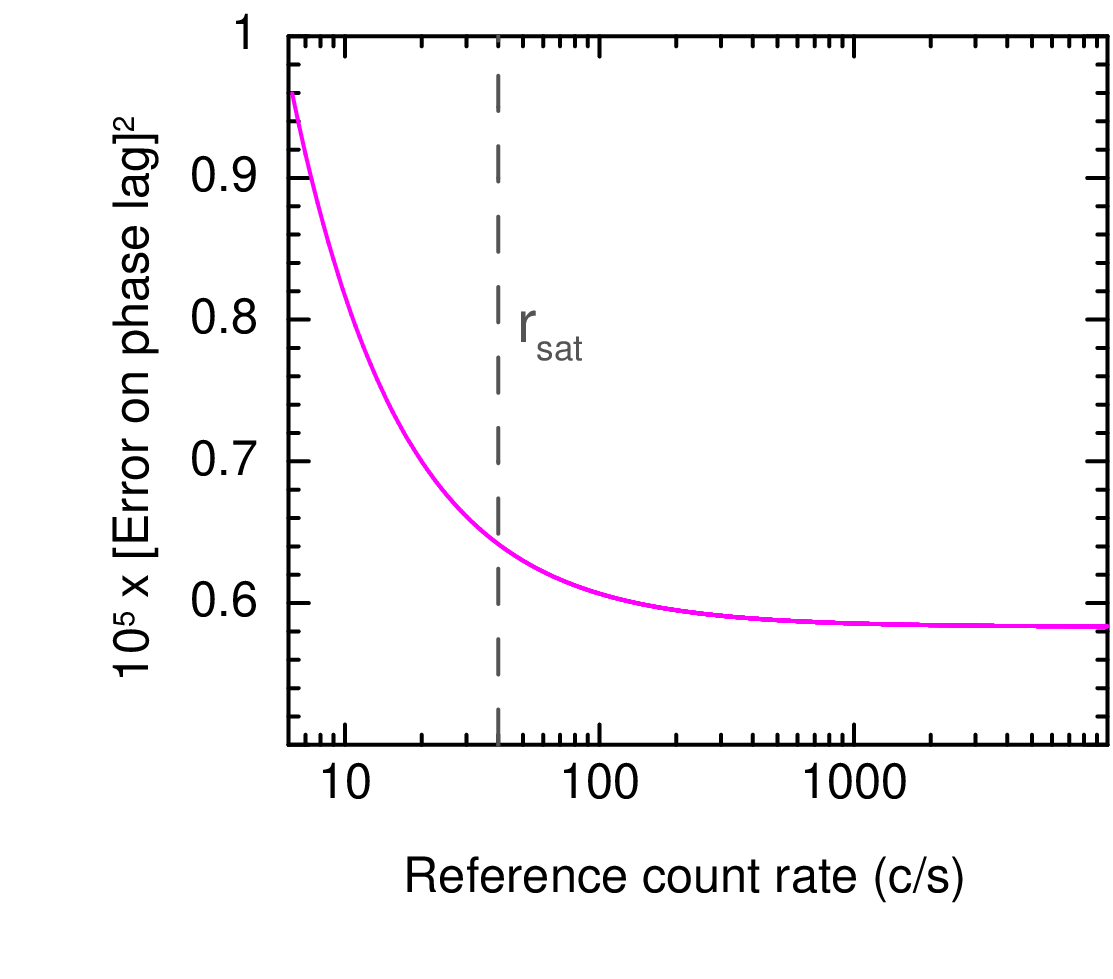}
\vspace{-5mm}
 \caption{Square of the 1 $\sigma$ error on the phase lag (in radians)
   between a $\psi$ bin with a count rate of 3 c/s and a reference
   time series, plotted against the count rate of the reference time
   series, $\langle r \rangle$. The magenta solid line depicts the
   total error and the other lines in the left hand plot show the 4
   separate terms in the formula for the error (see Equation
   \ref{eqn:fracerr}). The right hand plot is a zoom in of the
   left. The grey dashed vertical line depicts the `saturation count
   rate' at which the error transitions from a very steep function of
   $\langle r \rangle$ to a more shallow dependence (see Equation
   \ref{eqn:sat}).}
 \label{fig:errors}
\end{figure*}

\section{The importance of a high count rate reference
time series}
\label{sec:errors}

It is clear from Fig. \ref{fig:X2} that the reference count rate
is important for detection. In this Section we explore the importance
of the reference time series count rate, to assess how much of an advantage
can be gained simply by using an instrument with higher effective area
than the polarimeter to collect the reference time series. We first
note that the most important property of the reference time series is
that it is highly correlated with the polarimeter light curve. This is
guaranteed if instruments with a similar spectral response are used to
measure the polarimeter and reference time series. Therefore,
instruments sensitive to the same $\sim 2-8$ keV band as the GPDs,
such as the \textit{Neutron star Interior Composition ExploreR}
(\textit{NICER}: \citealt{Gendreau2016}), the X-Ray Concentrator Array
(XRCA) of \textit{STROBE-X} and the Spectroscopic Focusing Array (SFA) of
\extp~will provide an advantage in this respect. In practice though,
hard and soft X-rays tend to be highly correlated for the case of
accreting compact objects (e.g. \citealt{Nowak1999}), and so the
harder response of the LAD is unlikely to be a problem. We therefore
assume unity coherence throughout this paper.

\subsection{The polarisation degree required for detection}

Fig \ref{fig:pdet} shows the mean polarisation degree required to
detect polarisation variability with $5 \sigma$ confidence plotted
against mean reference time series count rate. Again, results for the
high and low inclination models are plotted on the left and right
respectively. To calculate the `detection polarisation degree',
$p_{\rm det}$, plotted on the y-axis, we find the $\langle p_0
\rangle$ value for which the $F$ statistic for the simulation
corresponds to $5 \sigma$ confidence. For example, the solid
black line in Fig \ref{fig:X2} (left), representing $\langle s \rangle
= \langle r \rangle =100$ c/s, crosses the $5 \sigma$ level
(upper dashed line) for $\langle p_0 \rangle \approx 5.5
\%$. Therefore, $p_{\rm det}\approx 5.5\%$ for \ixpe~($\langle s
\rangle=100$ c/s) when the reference time series mean count rate is
$\langle r \rangle =100$ c/s.

We show results for three different polarimeters: \ixpe~(black),
\xipe~(which is the same as the baseline \extp~polarimeter; red) and
\extp~goal (blue). For each line, we only consider $\langle r \rangle
\geq \langle s \rangle$, since in practice there is no need to use a
reference time series with a lower mean count rate than can be
provided by the polarimeter itself. We again assume a $200$ ks
exposure. We see that increasing the area of the polarimeter (from
\ixpe~to \xipe~to \extp~goal) has a large  impact on
sensitivity. Increasing the reference count rate also has a
significant effect up until $\sim 5000$ c/s where the sensitivity
starts to saturate. This is interesting, since \astrosat~can achieve
count rates of $\sim 5000$ c/s for a bright source as specified at the
start of Section \ref{sec:sims}. Therefore, for the simulation
parameters considered, \astrosat~would perform comparably to the
\extp~or \textit{STROBE-X} versions of the LAD over a 200 ks exposure
(that is, if the coherence between the LAD and GPD energy bands is
high as assumed, otherwise \textit{NICER} or the XRCA of
\textit{STROBE-X} may provide an advantage). We show below, however,
that there are other parameter combinations for which the LAD gives a
large advantage. It is also important to note that a $200$ ks exposure
on a low Earth orbit satellite will take $\sim 5$ days to collect,
over which time the QPO frequency will change fairly
significantly. This would therefore need to be accounted for in the
analysis [e.g. using techniques similar to the \cite{Mendez1998}
`shift and add' technique employed for kHz QPOs, or the
\cite{Tomsick2001} `stretch and bin' technique employed for LF
QPOs]. Finally, the detection polarisation degree is related very
simply to the modulation factor, $\mu$. Doubling $\mu$ would half
$p_{\rm det}$, and so polarimeter designs with larger $\mu$ are
unsurprisingly more sensitive.

\subsection{The `saturation count rate'}

We can further understand the role of the reference time series by
exploring the error on the phase lag. From Equation \ref{eqn:error},
the squared error is
\begin{equation}
\left[ d\phi(\psi_i,\nu) \right]^2 \propto
1 + \frac{P_{\rm noise}(\psi_i)}{P(\psi_i,\nu)} + \frac{P_{\rm noise} }{ P(\nu)} +
\frac{P_{\rm noise} P_{\rm noise}(\psi_i) }{ P(\nu) P(\psi_i,\nu) }.
\label{eqn:fracerr}
\end{equation}
Here, $P(\nu)$ and $P(\psi_i\nu)$ are respectively the intrinsic
(i.e. no Poisson noise) power spectra of the reference time series and
the light curve for the $i^{\rm th}$ $\psi$ bin. $P_{\rm noise}$ and
$P_{\rm noise}(\psi_i)$ are respectively the Poisson noise
contribution for the reference time series and the light curve for the
$i^{\rm th}$ $\psi$ bin. Fig. \ref{fig:errors} (left) shows this
squared error for a $\psi$ bin with mean count rate $3$ c/s, as a
function of reference band count rate, $\langle r \rangle$. We assume
a QPO with $10\%$ fractional rms (consistent with the high inclination
model), $T=200$ ks and $\Delta=0.2$ Hz. The solid magenta line shows
the total, whereas the black dashed line, the red dot-dashed line, the
green dotted line and the blue triple dot dashed line depict
respectively the first, second, third and fourth terms on the right
hand side of Equation \ref{eqn:fracerr}. We see that the first two
terms do not depend on $\langle r \rangle$, whereas the third and
fourth terms reduce with $\langle r \rangle$. We also see that the
second and fourth terms dominate over the others for this example,
which turns out to be generally the case for the set of observed LF
QPO properties.

Therefore, in the regime in which the fourth term dominates, we gain
an enormous advantage by increasing the area used to collect our
reference time series. If instead the second term dominates, any
further increase in $\langle r \rangle$ gives a more incremental
improvement. We can estimate this `saturation count rate' by setting
the second and fourth terms equal to one another to obtain
\begin{equation}
r_{\rm sat} = \frac{2(1+\langle b \rangle / \langle s \rangle)\Delta}{
  {\rm rms}^2 },
\label{eqn:sat}
\end{equation}
where $\langle b \rangle / \langle s \rangle$ is the fractional
contribution of the polarimeter background (in this discussion we only
consider $b=0$). We see that $r_{\rm sat}=40$ c/s for the parameters used in
Fig. \ref{fig:errors}. It is important to note that $r_{\rm sat}$
marks only a change in regime. It is still possible to get an
improvement in signal to noise by increasing $\langle r \rangle$, even
for $\langle r \rangle >r_{\rm sat}$. Fig \ref{fig:errors} (right) - a
zoom in of Fig \ref{fig:errors} (left) - demonstrates this. The error
is still decreasing for $\langle r \rangle >r_{\rm sat}$ (grey dashed
line), until $\langle r \rangle >>r_{\rm sat}$. We also see this in
Fig \ref{fig:pdet}, where the sensitivity still improves until
$\langle r \rangle \sim 5000$ c/s, even though the saturation count
rate is $r_{\rm sat}=40$ c/s.

Nonetheless, Equation \ref{eqn:sat} shows that a high count rate
reference time series is most important for low rms, high $\Delta$
QPOs. Since LF QPOs have a roughly constant quality factor of $Q\sim
8$, high $\Delta$ translates to high $\nu_{qpo}$. Therefore, if we
wish to observe QPOs with \ixpe~alone, we should target the lowest
frequency, highest rms QPOs. For a fractional rms of $10\%$, we
calculate that this transition in regime occurs at $\nu_{ qpo} \sim 4$
Hz (by setting $r_{\rm sat}=100$ c/s, since this is the count rate
achievable by \ixpe~alone).

For $\nu_{qpo} \gtrsim 4$ Hz a high count rate reference time series
will vastly improve signal to noise, much more so than shown in Fig.
\ref{fig:pdet}. We do however need to keep in mind time lost to lining
up orbits of two observatories, except for \extp, which has both the
polarimeter and the LAD onboard. It is clear from our discussion in
this section that the biggest advantage afforded by a high count rate
reference time series is for high frequency (HF) QPOs, or indeed kHz
QPOs in NSs, which may display modulations in polarisation
properties if they are due to e.g. orbiting hot spots on the disc
(\citealt{Beheshtipour2016}). Taking the upper HF QPO from the triplet
of QPOs measured in GRO J1655-40 by \cite{Motta2014} ($\Delta=30$ Hz,
rms=$4.5\%$), gives $r_{\rm sat} \approx 29,630$ c/s. This count rate
is not achievable with current instrumentation, but the count rate for
both the \textit{STROBE-X} and \extp~versions of the LAD will be
higher than this for a bright source. These next-generation detectors
will therefore provide an enormous advantage over current
instrumentation.

\section{Discussion \& Conclusions}
\label{sec:conc}

We present a simple and robust method for detecting fast stochastic
variability in the polarisation properties of an X-ray source. Whereas
coherent oscillations can be detected using phase-folding, this is not
the case for broad band noise or even QPOs - for which the phase of
the oscillation does not increase predictably with time. We
demonstrate that, with our method, ruling out constant polarisation
degree and angle is simple for any kind of stochastic variability. If
we see sinusoidal modulations in the fractional rms and phase as a
function of modulation angle $\psi$ for any frequency range, we can
conclude that there is variability in the polarisation
properties. We also introduce a method to measure the amplitude and
phase of the individual oscillations in polarisation degree and angle
for the case of QPOs, which is analogous to the spectral-timing method
used in \cite{Ingram2016,Ingram2017}. However, also taking this extra
step for the case of broad band variability is more difficult. This is
because the bi-spectrum of the variability may mean that contamination
from other frequencies becomes important (\citealt{Kim1979}). QPOs
present a special limit of the bi-spectrum for which the phases of
different QPO harmonics are strongly coupled to one another
(\citealt{Ingram2015}; or in other words the bi-coherence is high
between different QPO harmonics: \citealt{Maccarone2011}), and the rms
near the centroid frequency of a given QPO harmonic is dominated by
that harmonic. We will consider broad band variability more closely in
a later work.

We note that the method explored here does not take account of the
phase difference between the QPO harmonics, since the phase of the
cross-spectrum is relative to a reference
time series. It is however simple to include this extra detail in the
analysis using the method of \cite{Ingram2015}, which will provide
further information on the physics of the system. We also note that it
is possible to use an equivalent method using Stokes parameters, which
we will explore in a future work.

We use our new method to investigate the detectability of the QPOs in
polarisation degree and angle predicted by the Lense-Thirring
precession model (\citealt{Ingram2015a}). We consider the predictions
for a high (more edge-on) inclination ($i=70^\circ$) and low
inclination ($i=30^\circ$) object, and vary the mean polarisation
degree of the source, which is critical for detectability and a key
model uncertainty. We find that \ixpe~will be able to detect the
oscillations in a $200$ ks exposure, providing the mean polarisation
degree is $\langle p_0 \rangle \gtrsim 5.5\%$ and $\langle p_0 \rangle
\gtrsim 2.6\%$ for the high and low inclination models
respectively. The difference between models results from the swings in
polarisation angle being greater for a precessing vector being viewed
from above as opposed to from the side. The mean polarisation degree
is however \textit{predicted} to be lower for low inclination objects,
so high inclination objects will still likely be better targets
(\citealt{Chandrasekhar1960,Sunyaev1985}). Utilising a simultaneous
exposure with a larger area detector such as \textit{AstroSat} or
\textit{NICER} reduces the required polarisation degree down to
$\langle p_0 \rangle \gtrsim 4.7\%$ and $\langle p_0 \rangle \gtrsim
1.7\%$ for high and low inclination objects respectively. This is
encouraging, since the existing \textit{OSO 8} polarisation
measurements for Cygnus X-1, a low inclination ($i \approx 30^\circ$;
\citealt{Orosz2011a}) source, are $\langle p_0 \rangle = 2.4 \pm 1.1
\%$ and $\langle p_0 \rangle = 5.3 \pm 2.5 \%$ at $\sim 2.6$ keV and
$\sim 5.2$ keV respectively (\citealt{Long1980}). It is debatable
whether or not Cygnus X-1 displays QPOs in its flux at all
(\citealt{Axelsson2013}; Rapisarda et al submitted), but if there is
indeed precession in this source, the apparent lack of QPOs could be
due to the high amplitude broadband variability dominating over a low
amplitude QPO (consistent with the low measured inclination angle). In
this case, we would expect to see QPOs in the polarisation properties
using \ixpe~and \astrosat. We also expect to detect these QPOs in the
higher inclination sources, if they do indeed have a  larger mean
polarisation degree than Cygnus X-1 as predicted.

Interestingly for the prospect of polarimetry-timing, recent general
relativistic magnetohydrodynamic simulations show that the jet is
expected to precess in step with a precessing accretion flow (Liska et
al in prep). If optically thin synchrotron emission from the jet also
contributes significantly to the X-ray flux for some states, we would
expect still a higher polarisation degree for such states
(e.g. \citealt{Rybicki1979}), and therefore would expect detection of
of polarisation variability to be easier than we estimate here. For the hard
state, a dominant jet contribution to the X-rays has been argued
against on the basis of e.g. X-ray/radio scaling relations and
energetics (\citealt{Maccarone2005,Malzac2009}). In addition,
\cite{Heil2015} find that higher inclination sources
have harder X-ray spectra, which is not expected for emission from an
outflowing jet. Indeed, a precessing jet cannot explain Type-C
QPOs, at least in the X-rays, since their amplitude increases with
inclination angle. The trend should be the opposite for a precessing
jet, which would produce QPOs with amplitude roughly $\propto [1-
(v/c) \cos i]^{-2}$, where $v$ is the speed of the outflow.

However, it is possible that jet emission becomes more important in
the soft intermediate state (SIMS) when the source starts to display
Type-B QPOs. The amplitude of Type-B QPOs is higher for \textit{lower}
inclination sources (\citealt{Motta2015}), and jet precession has
previously been suggested as an origin (\citealt{Stevens2016}). Such a
switch in the X-ray
luminosity of the jet predicts that the ratio of power-law flux to
disk flux in the SIMs should be higher for lower inclination objects
due to beaming. This does not appear to be the case in the data (see
Fig 4 in \citealt{Gao2017}), in fact there are even hints of the
opposite trend, although the number of data points are too few to be
conclusive. Still, the relative importance of optically thin
synchrotron emission in the X-rays will be easy to test with \ixpe,
through simply measuring the polarisation degree. Jet precession has
also been suggested to explain observed infrared QPOs
(\citealt{Kalamkar2016}), which would predict a QPO in the infrared
polarisation angle. Finally, \textit{INTEGRAL} observations suggest
that the $\gamma$-ray ($0.4-2$ MeV) emission from Cygnus X-1 is highly
polarised (\citealt{Laurent2011}). It would be very interesting to
search for variability in this polarised $\gamma$-ray emission, but
the count rate achievable with \textit{INTEGRAL} ($\sim 0.03$ c/s) is
too low.


Although simultaneous observation with a large area detector improves
statistics, we caution that in practice much time will be lost to
lining up the orbits of two satellites, particularly for low Earth
orbits. A huge advantage will therefore be gained by use of a
satellite such as \extp, which is proposed to have a large area
detector and a polarimeter in the same payload. Higher signal to noise
can of course be achieved with longer exposure times, although the
drift in QPO frequency over timescales of $\sim$days will need to be
taken into account. We also show that the impact of using a large area
to collect the reference time series is maximised for higher
frequency QPOs. In particular, the study of polarisation in HF QPOs
will be inaccessible to \ixpe~but may be possible if a very large area
detector such as the LAD on \extp/\textit{STROBE-X} is used to collect
the reference time series.

The source code used to make some of the plots in this paper can be
downloaded from
\url{https://bitbucket.org/adingram/polarimetry-timing}. At the time
of writing, the repository contains the code used to create
Figs. \ref{fig:i70} and \ref{fig:i30} from this paper. In addition, we
plan to further develop this repository over time, with the ambition
of developing public software compatible with the eventual
\ixpe~pipeline.

\section*{Acknowledgements}

A. I. acknowledges support from the Netherlands Organisation for
Scientific Research (NWO) Veni Fellowship, grant number
639.041.437.


\bibliographystyle{/Users/adamingram/Dropbox/bibmaster/mnras}
\bibliography{/Users/adamingram/Dropbox/bibmaster/biblio}

%
\appendix

\section{Calculation of rms and phase lags}
\label{sec:bbn}
%
%

In this Appendix, we present the details of our calculation to arrive
at rms and phase lag as a function of $\psi$ from the plots of flux,
polarisation degree and polarisation angle presented in Figs. 6 and 8
of \cite{Ingram2015a}. We start from the cross-spectrum as defined by
Equation \ref{eqn:cross} in the main text. From this, it is useful to
define the \textit{complex covariance} (Mastroserio, Ingram \& van der
Klis in prep)
\begin{equation}
G(\psi_i,\nu,\Delta) = \frac{ C(\psi_i,\nu) }
    { \sqrt{P(\nu)} } \sqrt{\Delta},
\label{eqn:cov}
\end{equation}
where $P(\nu)$ is the intrinsic (i.e. white noise subtracted) power
spectrum of the reference time series. We use absolute rms
normalisation throughout (see \citealt{Ingram2013}
for a discussion on normalisation), such that the amplitude of the
complex covariance is the \textit{covariance}, as defined by
e.g. \cite{Wilkinson2009}; \cite{Uttley2014}. For unity coherence, the
covariance is equal to the absolute rms amplitude, but the statistical
uncertainties are smaller than those associated with calculating the
rms directly (\citealt{Wilkinson2009}). The complex covariance is
therefore a very useful statistic, since its amplitude gives absolute
rms as a function of $\psi$ and its argument gives phase lag with
respect to the reference time series as a function of $\psi$.

We now consider how quasi-periodic modulations in $p_0$ and $\psi_0$
affect the rms and phase lags as a function of modulation angle,
$\psi$. The count rate in the $i^{\rm  th}$ $\psi$ bin as a function
of QPO phase, $\omega$, is given by
\begin{equation}
s(\psi_i,\omega) = s(\omega)
f(\psi_i|\psi_0(\omega),p_0(\omega),\mu) \Delta\psi_i
\label{eqn:qpomod}
\end{equation}
We use the \cite{Ingram2015a} calculations for the functions
$s(\omega)$, $p_0(\omega)$ and $\psi_0(\omega)$. For each $\psi$ bin,
we evaluate $s(\psi_i,\omega)$ using the above equation for 32 QPO phases and take
the FT. Since the model in Equation (\ref{eqn:qpomod}) is periodic,
its FT can be represented as $S(\psi_i,k)$, where $k$ represents the
$k^{\rm th}$ harmonic. The amplitude of $S(\psi_i,k)$ gives the total
rms in the $k^{\rm th}$ harmonic for the $i^{\rm th}$ $\psi$
bin. Therefore, this relates to the FT of a
\textit{quasi-periodic} function, $S(\psi_i,\nu_k)$, as
$S(\psi_i,k)=S(\psi_i,\nu_k)\sqrt{\Delta_k}$, where $\nu_k$ and
$\Delta_k$ are respectively the centroid and FWHM of the $k^{\rm th}$
QPO harmonic. We assume that the reference time series is related to
the polarimeter count rate as $r(t) = s(t) \langle r \rangle / \langle
s \rangle$. This assumption simplifies our calculations, and any
deviation from this (due e.g. to the detector  used to measure the
reference time series having a different instrument response to that
of the polarimeter) will not affect our conclusions at
all. Substituting this into Equation (\ref{eqn:cov}), our model for
the complex covariance at the $k^{\rm th}$ QPO harmonic becomes
\begin{equation}
G(\psi_i,\nu_k) = S(\psi_i,k) \exp[ - i \phi_r(k) ],
\end{equation}
where $\phi_r(k) = \arg\{ S(k) \}$ is the phase of the $k^{\rm th}$
harmonic of the reference time series.

Even though the QPO tends to dominate the variability amplitude at the
centroid frequency of the fundamental, there is still a contribution
to the variability in the flux from the broad band noise (BBN). This is
noise in the sense that it is stochastic and has no favoured
characteristic frequency, but it is intrinsic to the source (i.e. not
instrumental). It is possible that there is BBN variability in the
polarisation properties as well as the flux, which would be very
interesting in itself. However, if the polarisation properties
associated with the BBN signal are intrinsically \textit{constant},
then the presence of the BBN will \textit{dilute} the observed
variability in polarisation properties driven by the QPO. It is
therefore prudent to take this into account.

Around the QPO fundamental frequency, the BBN is characterised by
approximately constant $\nu P(\nu)$. For $P(\nu)$ in units of squared
fractional rms per Hz, this constant level is $\sim 0.01$. The squared
fractional rms of the BBN in the frequency range of the $k^{\rm th}$
QPO harmonic, ${\rm rms}_n$, is equal to the integral of the BBN power in the range
$\nu_{k}-\Delta_k/2$ to $\nu_{k}+\Delta_k/2$. This gives ${\rm rms}_n^2
\sim 0.01 / Q$, and therefore ${\rm rms}_n  \sim 3\%$ (assuming
$Q=8$). We use this value for the BBN
throughout. We can then imagine adding the QPO and BBN signals
together to get the total reference time series. In Fourier space,
this is simply
\begin{equation}
R_{\rm tot}(\nu) = R(\nu) + R_n(\nu),
\end{equation}
with a similar expression for the light curve corresponding to each
$\psi$ bin of the polarimeter
\begin{eqnarray}
S_{\rm tot}(\psi_i,\nu) &=& S(\psi_i,\nu) + S_n(\psi_i,\nu) \nonumber \\
&=& S(\psi_i,\nu) + s(\psi_i) R_n(\nu) / \langle r \rangle.
\label{eqn:spsit}
\end{eqnarray}
Here, subscript $n$ refers to the BBN component and $s(\psi_i)$ is the
time-averaged count rate for the $i^{\rm th}$ $\psi$ bin. For the
above equation, we have assumed that: 1) there is no variability in
polarisation properties of the BBN, and 2) the polarimeter and
reference signals are the same as one another except for their mean
count rate. Assumption (1) is the most prudent assumption we can make,
since it allows us to test the possibility that modulations caused by
the QPO are diluted by the BBN. Assumption (2) simplifies the
expressions without making any material difference.

If the QPO and BBN are not correlated with one another, the complex
covariance becomes
\begin{equation}
G_{\rm tot}(\psi_i,\nu) = \frac{\langle S(\psi_i,\nu) R^*(\nu) \rangle + \langle S_n(\psi_i,\nu) R_n^*(\nu) \rangle}{
  \sqrt{P(\nu)} } \sqrt{\Delta}.
\end{equation}
Using Equation (\ref{eqn:spsit}) and rearranging gives
\begin{equation}
G_{\rm tot}(\psi_i,\nu) = \frac{\Delta \langle S(\psi_i,\nu) R^*(\nu)/
  \langle r \rangle \rangle + \Delta \langle s(\psi_i)
  |R_n(\nu)|^2/\langle r \rangle^2 \rangle}{ \sqrt{P(\nu)
    \Delta/\langle r \rangle^2} }.
\end{equation}
Assuming unity coherence and tidying up by using fractional rms leaves
us with
\begin{equation}
G_{\rm tot}(\psi_i,\nu_k) = \frac{ {\rm rms}_q(k) S(\psi_i,k) \exp[ - i
  \phi_r(k) ] + {\rm rms}_n^2 s(\psi_i) } { \sqrt{ {\rm rms}_q^2(k)
  + {\rm rms}_n^2} },
\label{eqn:gtot}
\end{equation}
for the total model complex covariance at the $k^{\rm th}$ QPO
harmonic. Here, ${\rm rms}_q(k) = |S(k)| \sqrt{\Delta} / \langle r
\rangle$. From Equation \ref{eqn:gtot}, we can easily calculate
fractional rms, $|G_{\rm tot}(\psi_i,\nu_k)|/s(\psi_i)$, and phase
lag, $\arg \{ G_{\rm tot}(\psi_i,\nu_k) \}$ for each QPO
harmonic. Throughout this paper, we use fractional rms and phase as
the diagnostics to be compared with
(synthetic / future) data. For the case of spectral-timing, a number
of reasons make it more statistically convenient to instead work with
real and imaginary parts of the complex covariance or cross-spectrum
(\citealt{Ingram2016,Rapisarda2016,Ingram2017}; Mastroserio, Ingram \&
van der Klis in prep). However, not all of these reasons translate to
polarimetry-timing, and rms and phase are generally more intuitive.

\section{Simulation details}
\label{sec:simdet}

The $1~\sigma$ error on the complex covariance can be written as
\begin{equation}
dG(\psi_i,\nu) = \sqrt{ \frac{ [ P(\nu)+P_{\rm noise} ] [ P(\psi_i,\nu) +
    P_{\rm noise}(\psi_i) ] } { 2 T P(\nu) }  },
\label{eqn:error}
\end{equation}
where $T$ is the exposure time in seconds and $P(\psi_i,\nu)$ and
$P_{\rm noise}(\psi_i)$ are respectively the intrinsic and Poisson
noise power-spectrum of the $i^{\rm th}$ modulation angle bin. This
comes from the expression for the error on the cross-spectrum
(\citealt{Vaughan1994}), with the number of realisations set to $T
\Delta$. In absolute rms normalisation, the Poisson noise level of the
reference time series is $P_{\rm noise} = 2 ( \langle r \rangle +
\langle b_r \rangle )$, where $\langle b_r \rangle$ is the mean
background count rate. The corresponding expression for $P_{\rm
  noise}(\psi_i)$ is similar
(e.g. \citealt{vanderKlis1989,Uttley2014}). We input as simulation
parameters $T$, $\Delta$, the total polarimeter count rate $\langle r
\rangle$ and $\langle s \rangle$. Everything else can be determined
from the model $G_{\rm tot}(\psi_i,\nu_k)$. We assume unity
coherence, making it simple to estimate the power in each $\psi$
bin. We then generate Gaussian random variables for the real and
imaginary parts of our simulated model $G_{\rm tot}(\psi_i,\nu_k)$,
and calculate the fractional amplitude and phase from that, using
standard error propagation.


\bsp	
\label{lastpage}
\end{document}